%% file: B2G-16-003_temp.tex
\pdfoutput=1

\documentclass[11pt,twoside,a4paper,cmspaper,final,collab]{cms-tdr}
\def\svnVersion{391257}\def\cmsCernNoTag{CERN-EP/2016-226}\def\cmsCernDate{\today}\def\cmsMessage{Published in Physics Letters B as \href{http://dx.doi.org/10.1016/j.physletb.2017.02.040}{\doi{10.1016/j.physletb.2017.02.040.}}}

\begin{document}\cmsNoteHeader{B2G-16-003}

\hyphenation{had-ron-i-za-tion}
\hyphenation{cal-or-i-me-ter}
\hyphenation{de-vices}
\RCS$Revision: 379113 $
\RCS$HeadURL: svn+ssh://svn.cern.ch/reps/tdr2/papers/B2G-16-003/trunk/B2G-16-003.tex $
\RCS$Id: B2G-16-003.tex 379113 2016-12-28 14:40:11Z alverson $
\newlength\cmsFigWidth
\ifthenelse{\boolean{cms@external}}{\setlength\cmsFigWidth{\columnwidth}}{\setlength\cmsFigWidth{0.75\textwidth}}
\ifthenelse{\boolean{cms@external}}{\providecommand{\cmsLeft}{top\xspace}}{\providecommand{\cmsLeft}{left\xspace}}
\ifthenelse{\boolean{cms@external}}{\providecommand{\cmsRight}{bottom\xspace}}{\providecommand{\cmsRight}{right\xspace}}

\newcommand{\X}{\ensuremath{\cmsSymbolFace{X}}\xspace}
\newcommand{\V}{\ensuremath{\cmsSymbolFace{V}}\xspace}
\newcommand{\h}{\ensuremath{\cmsSymbolFace{H}}\xspace}
\renewcommand{\PW}{\ensuremath{\cmsSymbolFace{W}}\xspace}
\newcommand{\htobb}{\ensuremath{\PH\to\bbbar}\xspace}
\newcommand{\VH}{\ensuremath{\V\PH}\xspace}
\newcommand{\VV}{\ensuremath{\V\V}\xspace}
\newcommand{\ST}{\ensuremath{\text{t+X}}\xspace}
\newcommand{\Vjets}{\ensuremath{\V\text{+jets}}\xspace}
\newcommand{\mVH}{\ensuremath{m_{\VH}}\xspace}
\newcommand{\mtVH}{\ensuremath{m_{\VH}^\mathrm{T}}\xspace}
\newcommand{\mX}{\ensuremath{m_{\X}}\xspace}
\newcommand{\mH}{\ensuremath{m_{\PH}}\xspace}
\newcommand{\mj}{\ensuremath{m_{\mathrm{j}}}\xspace}
\newcommand{\B}{\ensuremath{\mathcal{B}}}
\newcommand{\gV}{\ensuremath{g_\text{V}}\xspace}
\newcommand{\cH}{\ensuremath{c_\text{H}}\xspace}
\newcommand{\cF}{\ensuremath{c_\text{F}}\xspace}
\newcommand{\MHT}{\ensuremath{H_\text{T}^\text{miss}}\xspace}
\newcommand{\MG}{\MADGRAPH{5}\_a\MCATNLO\xspace}

\cmsNoteHeader{B2G-16-003}
\title{Search for heavy resonances decaying into a vector boson and a Higgs boson in final states with charged leptons, neutrinos, and b quarks}

\date{\today}

\abstract{
  A search for heavy resonances decaying to a Higgs boson and a vector boson is presented. The analysis is performed using data samples collected in 2015 by the CMS experiment at the LHC in proton-proton collisions at a center-of-mass energy of 13\TeV, corresponding to integrated luminosities of 2.2--2.5\fbinv. The search is performed in channels in which the vector boson decays into leptonic final states ($\Z \to \nu\nu$, $\PW\to \ell \nu$, and $\Z \to \ell \ell$, with $\ell = \Pe,\Pgm$), while the Higgs boson decays to collimated b quark pairs detected as a single massive jet. The discriminating power of a jet mass requirement and a b jet tagging algorithm are exploited to suppress the standard model backgrounds. The event yields observed in data are consistent with the background expectation. In the context of a theoretical model with a heavy vector triplet, a resonance with mass less than 2\TeV is excluded at 95\% confidence level. The results are also interpreted in terms of limits on the parameters of the model, improving on the reach of previous searches.
}

\hypersetup{%
pdfauthor={CMS Collaboration},%
pdftitle={Search for heavy resonances decaying into a vector boson and a Higgs boson in final states with charged leptons, neutrinos, and b quarks},%
pdfsubject={CMS},%
pdfkeywords={CMS, physics, B2G, diboson, VH, semileptonic}
}

\maketitle

\section{Introduction}

The discovery of a Higgs boson \h at the CERN LHC~\cite{bib:Aad20121,bib:Chatrchyan201230,Chatrchyan:2013lba} suggests that the standard model (SM) mechanism that connects electroweak (EW) symmetry breaking to the generation of particle masses is largely correct.
However, the relatively light value of the Higgs boson mass $\mH = 125.09 \pm 0.21\stat \pm 0.11\syst\GeV$~\cite{Aad:2014aba,Khachatryan:2014jba,Chatrchyan:2014vua,Aad:2015zhl} leaves the hierarchy problem unsolved~\cite{1988NuPhB}, pointing to phenomena beyond the SM, which could be unveiled by searches at the LHC.
Many theories that incorporate phenomena beyond the SM postulate the existence of new heavy resonances coupled to the SM bosons. Among them, weakly coupled spin-1 $\PZpr$~\cite{Barger:1980ix,1126-6708-2009-11-068} and $\PW'$ models~\cite{Grojean:2011vu} or strongly coupled Composite Higgs~\cite{Contino2011,Marzocca2012,Bellazzini:2014yua}, and Little Higgs models~\cite{Han:2003wu,Schmaltz,Perelstein2007247} have been widely discussed.

A large number of models are generalized in the heavy vector triplet (HVT) framework~\cite{Pappadopulo2014}, which introduces one neutral ($\PZpr$) and two electrically charged ($\PW'$) heavy resonances.
The HVT model is parametrized in terms of three parameters: the strength  $\gV$ of a new interaction; the coupling $\cH$ between the heavy vector bosons, the Higgs boson, and longitudinally polarized SM vector bosons; and the coupling $\cF$ between the HVT bosons and the SM fermions.
In the HVT scenario, model~B with parameters $\gV = 3$, $\cH = 0.976$, and $\cF = 1.024$~\cite{Pappadopulo2014} is used as the benchmark. With these values, the couplings of the heavy resonances to fermions and to SM bosons are similar, yielding a sizable branching fraction for the heavy resonance decay into a SM vector boson \PW or \Z (generically labeled as \V) and a Higgs boson~\cite{Pappadopulo2014}.

Bounds from previous searches~\cite{Khachatryan:2015lba,Khachatryan:2015bma,Khachatryan:2015ywa,Khachatryan:2016yji} require the masses of these resonances to be above 1\TeV in the HVT framework. In this mass region, the two bosons produced in the resonance decay would have large Lorentz boosts in the laboratory frame. When decaying, each boson would generate a pair of collimated particles, a distinctive signature, which can be well identified in the CMS experiment.
Because of the large predicted branching fraction, the decay of high-momentum Higgs bosons to $\bbbar$ final states is considered. The Higgs boson is reconstructed as one unresolved jet, tagged as containing at least one bottom quark. Backgrounds from single quark and gluon jets are reduced by a jet mass requirement. In order to discriminate against the large multijet background, the search is focused on the leptonic decays of the vector bosons ($\Z \to \nu\nu$, $\PW \to \ell \nu$, and $\Z \to \ell \ell$, with $\ell = \Pe,\mu$).

The main SM background process is the production of vector bosons with additional hadronic jets (\Vjets). The estimation of this background is based on events in signal-depleted jet mass sidebands, with a transfer function, derived from simulation, from the sidebands to the signal-enriched region.
Top quark production also accounts for a sizable contribution to the background in $1\ell$ final states, and is determined from simulation normalized to data in dedicated control regions. Diboson production processes, including pairs of vector bosons (\VV) and the SM production of a Higgs boson and vector boson (\VH), represent minor contributions to the overall background and are estimated from simulation.
A signal would produce a localized excess above a smoothly falling background in the distribution of the kinematic variable \mVH, whose definition and relationship to the resonance mass \mX depends on the final state. Results are interpreted in the context of HVT models in the benchmark scenario~B~\cite{Pappadopulo2014}.

\section{Data and simulated samples}\label{sec:mcsimulation}

The data samples analyzed in this study were collected with the CMS detector in proton-proton collisions at a center-of-mass energy of 13\TeV during 2015.
The samples correspond to integrated luminosities of 2.2--2.5 \fbinv, depending on the final state considered.

Simulated signal events are generated at leading order (LO) according to the HVT model~B~\cite{Pappadopulo2014} with the \MG v5.2.2.2 matrix element generator~\cite{MadGraph}. The Higgs boson is required to decay into a \bbbar pair, and the vector boson into leptons. A contribution from vector boson decays into $\tau$ leptons is also included through subsequent decays to $\Pe$ or $\mu$ that satisfy the event selection.
Different \mX hypotheses in the range 800 to 4000\GeV are considered, assuming a resonance width narrow enough (0.1\% of the resonance mass) to be negligible with respect to the experimental resolution. This approximation is valid in a large fraction of the HVT parameter space, and will be discussed in Section~\ref{sec:res}.

The analysis utilizes a set of simulated samples to characterize the main SM background processes.  Samples of \Vjets events are produced with \MG and normalized to the next-to-next-to-leading-order (NNLO) cross section, computed using {\FEWZ} v3.1~\cite{Li:2012wna}. The \V boson \pt spectra are corrected to account for next-to-leading-order (NLO) QCD and EW contributions~\cite{Kallweit:2015dum}. Top quark pair production is simulated with the NLO \POWHEG v2 generator~\cite{Nason:2004rx,Frixione:2007vw,Alioli:2010xd} and rescaled to the cross section value computed with \textsc{Top++} v2.0~\cite{Czakon:2011xx} at NNLO. Minor SM backgrounds, such as \VV and \VH production, and single top quark (\ST) production in $s$-channel, $t$-channel, and in t\PW associated production, are simulated at NLO with \MG. Multijet production is simulated at leading order with the same generator.

Parton showering and hadronization processes are simulated by interfacing the event generators to~\PYTHIA 8.205~\cite{Sjostrand:2007gs,Sjostrand:2006za} with the CUETP8M1~\cite{Skands:2014pea,Khachatryan:2015pea} tune. The NNPDF 3.0~\cite{Ball:2014uwa} parton distribution functions (PDFs) are used to model the momentum distribution of the colliding partons inside the protons. Generated events, including additional proton-proton interactions within the same bunch crossing (pileup) at the level observed during 2015 data taking, are processed through a full detector simulation based on {\GEANTfour}~\cite{Agostinelli:2002hh} and reconstructed with the same algorithms used for data.

\section{CMS detector}\label{sec:detector}

The central feature of the CMS detector is a superconducting solenoid of 6\unit{m} internal diameter. Within the solenoid volume are a silicon pixel and strip tracker, a lead tungstate crystal electromagnetic calorimeter (ECAL), and a brass and scintillator hadron calorimeter (HCAL), each composed of a barrel and two endcap sections. Forward calorimeters extend the pseudorapidity~\cite{Chatrchyan:2008zzk} coverage provided by the barrel and endcap detectors. Muons are measured in gas-ionization detectors embedded in the steel flux-return yoke outside the solenoid.

The silicon tracker measures charged particles within the pseudorapidity range $\abs{\eta}< 2.5$. It consists of 1440 silicon pixel and 15\,148 silicon strip detector modules and is located in the 3.8\unit{T} field of the solenoid. For nonisolated particles of transverse momentum $1 < \pt < 10\GeV$ and $\abs{\eta} < 1.4$, the track resolutions are typically 1.5\% in \pt and 25--90 (45--150)\mum in the transverse (longitudinal) impact parameter \cite{Chatrchyan:2014fea}.
The ECAL provides coverage up to $\abs{\eta} < 3.0$. The dielectron mass resolution for $\Z \to\Pe \Pe$ decays when both electrons are in the ECAL barrel is 1.9\%, and is 2.9\% when both electrons are in the endcaps.
The HCAL covers the range of $\abs{\eta} < 3.0$, which is extended to $\abs{\eta} < 5.2$ through forward calorimetry.
Muons are measured in the pseudorapidity range $\abs{\eta}< 2.4$, with detection planes made using three technologies: drift tubes, cathode strip chambers, and resistive-plate chambers. Combining muon tracks with matching tracks measured in the silicon tracker results in a \pt resolution of 2--10\% for muons with $0.1 < \pt < 1\TeV$~\cite{Chatrchyan:2012xi}.

The first level (L1) of the CMS trigger system, composed of custom hardware processors, uses information from the calorimeters and muon detectors to select the most interesting events in a fixed time interval of less than 4\mus. The high-level trigger (HLT) processor farm further decreases the event rate from around 100\unit{kHz} to about 1\unit{kHz}, before data storage.

A detailed description of the CMS detector, together with a definition of the coordinate system used and the relevant kinematic variables, can be found in Ref.~\cite{Chatrchyan:2008zzk}.

\section{Event reconstruction}\label{sec:eventreconstruction}

In CMS, a global event reconstruction is performed using a particle-flow (PF) algorithm~\cite{CMS-PAS-PFT-09-001,CMS-PAS-PFT-10-001}, which uses an optimized combination of information from the various elements of the CMS detector to reconstruct and identify individual particles produced in each collision. The algorithm identifies each reconstructed particle either as an electron, a muon, a photon, a charged hadron, or a neutral hadron.

The PF candidates are clustered into jets using the anti-\kt algorithm~\cite{Cacciari:2008gp} with a distance parameter $R=0.4$ (AK4 jets) or $R=0.8$ (AK8 jets).
In order to suppress the contamination from pileup, charged particles not originating from the primary vertex, taken to be the one with the highest sum of $\pt^2$ over its constituent tracks, are discarded. The residual contamination removed is proportional to the event energy density and the jet area estimated using the {\FASTJET} package~\cite{Cacciari:2011ma,Cacciari:2008gn}. Jet energy corrections, extracted from simulation and data in multijet, $\gamma$+jets, and Z+jets events, are applied as functions of the transverse momentum and pseudorapidity to correct the jet response and to account for residual differences between data and simulation. The jet energy resolution amounts typically to 5\% at 1\TeV~\cite{Khachatryan:2016kdb}. Jets are required to pass an identification criterion, based on the jet composition in terms of the different classes of PF candidates, in order to remove spurious jets arising from detector noise.
The pruning algorithm~\cite{Ellis:2009me}, which is designed to remove contributions from soft radiation and additional interactions, is applied to AK8 jets. The pruned jet mass \mj is defined as the invariant mass associated with the four-momentum of the pruned jet, after the application of the jet energy corrections~\cite{Khachatryan:2016kdb}.
The AK8 jets are split into two subjets using the soft drop algorithm~\cite{Dasgupta:2013ihk,Larkoski:2014wba}.

The combined secondary vertex algorithm~\cite{CMS-PAS-BTV-15-001} is used for the identification of jets that originate from b quarks (b tagging).
The algorithm uses the tracks and secondary vertices associated with AK4 jets or AK8 subjets as inputs to a neural network to produce a discriminator with values between 0 and 1, with higher values indicating a higher b quark jet probability.
The loose and the tight operating points are about 85 and 50\% efficient, respectively, for b jets with \pt of about 100\GeV, with a false-positive rate for light-flavor jets of about 10 and 0.1\%.

The missing transverse momentum vector \ptvecmiss is defined as the projection of the negative vectorial sum of the momenta of all PF candidates onto the plane perpendicular to the beams, and its magnitude is referred to as \MET. The missing hadronic activity \MHT is defined as the magnitude of the negative vectorial sum of the transverse momenta of all AK4 jets with $\pt > 20\GeV$.
Corrections for the \MET detector response and resolution are derived from $\gamma$+jets and Z+jets events, and applied to simulated events~\cite{CMS:2016ljj}.

Electrons are reconstructed in the fiducial region $\abs{\eta}<2.5$ by matching the energy deposits in the ECAL with tracks reconstructed in the tracker~\cite{Khachatryan:2015hwa}. The electron identification is based on the distribution of energy deposited along the electron trajectory, the direction and momentum of the track in the inner tracker, and its compatibility with the primary vertex of the event.  Additional requirements are applied to remove electrons produced by photon conversions. Electrons are further required to be isolated from other activity in the detector. The electron isolation parameter is defined as the sum of transverse momenta of all the PF candidates (excluding the electron itself) within $\Delta R = \sqrt{\smash[b]{(\Delta\eta)^2+(\Delta\phi)^2}} < 0.3$ around the electron direction, after the contributions from pileup and other reconstructed electrons are removed.
Photons are reconstructed as energy clusters in the ECAL, and are distinguished from jets and electrons using information that includes isolation and the transverse shape of the ECAL energy deposit.

Muons are reconstructed within the acceptance of the CMS muon systems, $\abs{\eta}<2.4$, using the information from both the muon spectrometer and the silicon tracker~\cite{Chatrchyan:2012xi}. Muon candidates are identified via selection criteria based on the compatibility of tracks reconstructed from silicon tracker information only with tracks reconstructed from the combination of the hits in both the tracker and muon detector.
Additional requirements are based on the compatibility of the trajectory with the primary vertex, and on the number of hits observed in the tracker and muon systems.
The muon isolation is computed from reconstructed tracks within a cone $\Delta R< 0.3$ around the muon direction, ignoring the muon itself.

Hadronically decaying $\tau$ leptons are reconstructed combining one or three hadronic charged PF candidates with up to two neutral pions, the latter also reconstructed by the PF algorithm from the photons arising from the $\Pgpz \to\gamma\gamma$ decay~\cite{Khachatryan:2015dfa}.

\section{Event selection}
\label{sec:evtsel}

The set of criteria used to identify the Higgs boson candidate is the same for each event category. The highest-\pt AK8 jet in the event is required to have $\pt>200\GeV$ and $\abs{\eta}<2.5$. The pruned jet mass \mj must satisfy $105 < \mj < 135\GeV$. The region $65 < \mj < 105\GeV$ is not used, to avoid overlaps with searches targeting resonant \VV final states. In order to discriminate against the copious vector boson production in association with light-flavored jets, events are classified according to the number of subjets (1 or 2) passing the loose b tagging selection; those failing this requirement are discarded.

Events are divided into categories depending on the number (0, 1, or 2) and flavor ($\Pe$ or $\mu$) of the reconstructed charged leptons, and the presence of either 1 or 2 b-tagged subjets in the AK8 jet.
The two categories with no charged leptons are referred to collectively as the zero-lepton ($0\ell$) channel. Similarly, the single-lepton ($1\ell$) and double-lepton ($2\ell$) channels each comprise four categories. In total, 10 exclusive categories are defined.

In the $0\ell$ channel, candidate signal events are expected to have a large \MET due to the boosted \Z boson decaying into a pair of neutrinos, which escape undetected. Data are collected using triggers that require \MET or \MHT greater than $90\GeV$, without including muons in the \MET or \MHT computation. A stringent selection is applied to the reconstructed \MET, which is required to be greater than 200\GeV, to ensure that the trigger is fully efficient.
The copious multijet production is greatly suppressed by imposing requirements on the minimum azimuthal angular separations between jets and the missing transverse momentum vector, $\Delta\phi\text{(jet, \ptvecmiss)}$. All the AK8 and AK4 jets in the event must satisfy $\Delta\phi\text{(jet, \ptvecmiss)} > 0.5$. The Higgs boson jet candidate must fulfill the tighter requirement $\Delta\phi\text{(jet, \ptvecmiss)} > 2$ and additional criteria designed to remove events arising from detector noise.
Events containing isolated leptons with $\pt>10\GeV$, hadronically-decaying $\tau$ leptons with $\pt>18\GeV$, and photons with $\pt>15\GeV$ are removed in order to reduce the contribution of other SM processes.
The \ttbar background contribution is reduced by removing events in which any AK4 jet, excluding the Higgs boson jet candidate, is b tagged using the loose operating point.
Because of the lack of visible decay products from the \Z boson, reconstruction of the resonance mass is not directly viable. Instead, the Higgs boson jet momentum and the \ptvecmiss are used to compute the transverse mass  $\mtVH = \sqrt{\smash[b]{2 \MET \ET^\text{jet}\, [1-\cos{\Delta \phi\text{(jet, \ptvecmiss)} }]}}$. This variable is utilized as an estimator of \mX for the $0\ell$ channel.

Events in the $1\ell$ channel are collected requiring one lepton to be reconstructed online. The \pt threshold at trigger level is 105\GeV for electrons and 45\GeV for muons. Offline, events are accepted if there is exactly one reconstructed electron or muon with \pt larger than 135\GeV or 55\GeV, respectively, passing restrictive selection criteria. Events with additional leptons passing looser selections, or hadronically decaying $\tau$ leptons, are discarded. In the single-electron channel, multijet background is reduced by requiring $\MET>80\GeV$. Azimuthal angular separations $\Delta \phi (\ell, \ptvecmiss) < 2$ and $\Delta \phi (\text{jet}, \ptvecmiss) > 2$ are required to select a back-to-back topology.
As for the $0\ell$ selection, events with additional b-tagged AK4 jets are vetoed.
The four-momentum of the \PW boson candidate is quantified using a kinematic reconstruction of the neutrino momentum.
The components of the neutrino momentum in the transverse plane are assumed to be equal to \ptvecmiss. By constraining the invariant mass of the charged lepton and neutrino to be equal to the \PW boson mass, a quadratic equation is derived for the longitudinal component of the neutrino momentum, $p_z^\nu$.
The reconstructed $p_z^\nu$ is chosen to be the real solution with the lower magnitude or, where both the solutions are complex, the real part with the lowest value.
If the \PW boson has a transverse momentum greater than 200\GeV, it is used to construct the resonance candidate mass \mVH, otherwise the event is discarded.

The $2\ell$ channel accepts events collected with the same triggers as in the $1\ell$ channel. An additional isolated electron or muon with $\pt > 20\GeV$, with the same flavor as the leading one and opposite charge, is required to be reconstructed and identified. In order to increase the signal efficiency, a looser identification requirement is applied to both electrons, and one of the two muons is allowed to be identified only in the tracker. If the isolation cones of the two muons overlap, the contribution of one is subtracted from the isolation calculation of the other in each case.
The \Z boson candidates are retained only if the dilepton invariant mass lies between $70$ and 110\GeV. The transverse momentum of the \Z boson candidate is required to be at least 200\GeV, otherwise the event is removed. Additionally, the separation in $\eta$ and $\phi$ between the \Z boson candidate and the Higgs boson jet is required to satisfy $| \Delta \eta (\Z,\text{jet})| < 5$ and $\Delta \phi (\Z,\text{jet}) >2.5$. Since the \ttbar contribution is small, no veto on additional b-tagged AK4 jets is applied.
The resonance candidate mass \mVH is defined as the invariant mass of the \Z boson and the AK8 jet.

The signal efficiency for the combined $0\ell$, $1\ell$, and $2\ell$ channels following these selections is 20--30\% for the 2 b-tagged subjet categories for a resonance mass $\mX = 1\TeV$, decreasing to about 10\% for $\mX = 4\TeV$. This reduction is due to the degradation of track reconstruction and b tagging performances at very large \pt, and to the smaller angle between the two b quarks, which tend to be reconstructed in one single subjet. The loss of efficiency is recovered by the 1 b-tagged subjet categories, which provide an additional 10\% signal efficiency at $\mX = 1\TeV$, and 20\% at $\mX = 4\TeV$.

\section{Estimated and observed background}
\label{sec:bkg}

The main source of background events originates from the production of a vector boson in association with jets, and the subsequent decay of the vector boson into one of the considered leptonic final states. This background is relevant both when genuine b jets are identified and when a jet originating from a lighter quark or a gluon is misidentified as originating from a b quark. In the $1\ell$ and $2\ell$ channels, the main contributions are due to $\PW \to \ell\nu$ and $\Z\to \ell\ell$ processes, respectively. In the $0\ell$ channel $\Z \to \nu\nu$ and $\PW \to \ell\nu$ processes account for approximately 60\% and 40\% of the \Vjets background, respectively. In the latter case, the lepton is either emitted outside the detector acceptance, or is not reconstructed and identified. A sizable background originates from b jets and \PW bosons from decays of pair-produced top quarks. Minor contributions come from \ST, \VV, \VH, and multijet processes.

The normalization of the top quark background (\ttbar and \ST) is determined in top quark enriched control regions where the simulated \mj and \mVH distributions are also checked against data. Four top quark control regions are defined, depending on the number of reconstructed leptons (0 or 1) and the number of b-tagged subjets (1 or 2). The top quark control regions are defined by inverting the b tagging veto on the AK4 jets in the event, and by applying a tight b tagging selection to obtain a \ttbar sample with higher purity. Data are found to be in agreement with the shape of the simulated \mj and \mVH distributions.
Multiplicative scale factors are derived for each region from the difference in normalization between data and simulation, after subtracting the contribution of the other backgrounds from the data. These factors, reported in Table~\ref{tab:TopCR}, are applied to correct the normalization of the \ttbar and \ST background. In the dilepton channel, due to the small number of events, the \ttbar normalization and shape are taken from simulation.

\begin{table}[!htb]
  \topcaption{Scale factors derived for the normalization of the estimated \ttbar and \ST backgrounds from simulation, for different event categories. Electron and muon categories are merged. Uncertainties due to the limited size of the event samples (stat) and the uncertainty in the b tagging efficiency (syst) are reported separately.}\label{tab:TopCR}
  \centering
    \begin{tabular}{cc cccc}
      \multicolumn{2}{c }{category} & scale factor & stat & syst \\
      \hline
      \multirow{2}{*}{1 b tag} & $1\ell$ & 0.82 & $\pm$0.03 & $\pm$0.04 \\
      & $0\ell$ & 0.85 & $\pm$0.06 & $\pm$0.04 \\
      \hline
      \multirow{2}{*}{2 b tag} & $1\ell$ & 0.83 & $\pm$0.07 & $\pm$0.04 \\
      & $0\ell$ & 0.54 & $\pm$0.13 & $\pm$0.02 \\
      \hline
    \end{tabular}
\end{table}

The contribution of the dominant \Vjets background is estimated through a procedure based on data.
Signal-depleted samples are defined, containing events that pass all selections described in Section~\ref{sec:evtsel} apart from the requirement on the pruned jet mass.
Two \mj sidebands (SB) are considered, and used to predict the background contributions in the signal region (SR). The lower and upper sidebands accept events falling in the ranges $30 < \mj < 65\GeV$ and $\mj > 135\GeV$, respectively.
Analytic functions are fitted to the distributions of \mj found in simulation, considering separately \Vjets, \ttbar and \ST, and all SM diboson production processes. The \mj spectrum in \Vjets events consists of a smoothly falling distribution, while diboson samples present one or two peaks corresponding to the \PW/\Z and Higgs boson masses. Top quark samples have instead one peak in the \mj spectrum for hadronically decaying \PW bosons and one for the top quark itself, in events where the hadronic \PW boson or top quark is reconstructed within the selected AK8 jet.

\begin{figure*}[!hbtp]\centering
\includegraphics[width=0.40\textwidth]{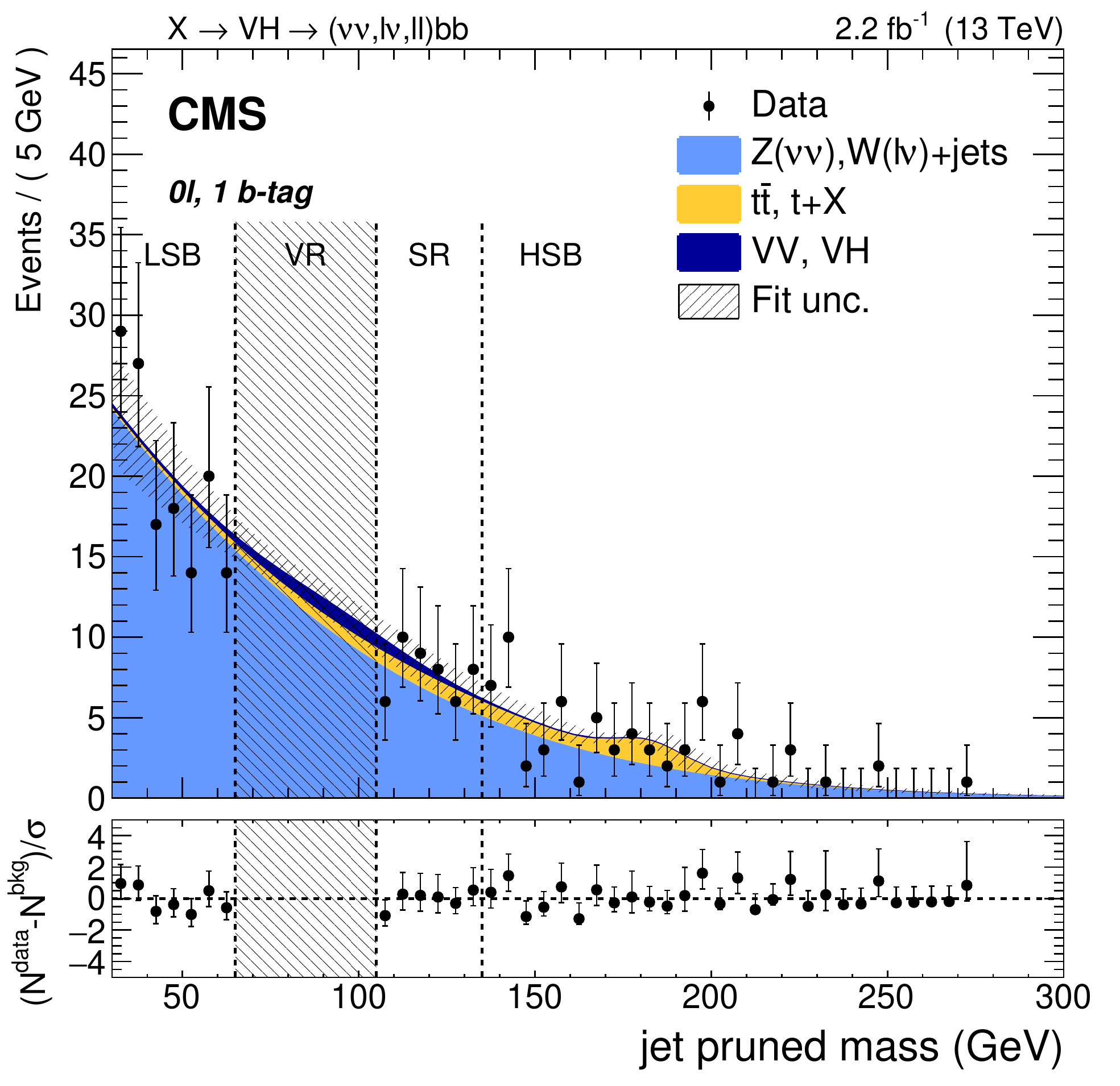}
\includegraphics[width=0.40\textwidth]{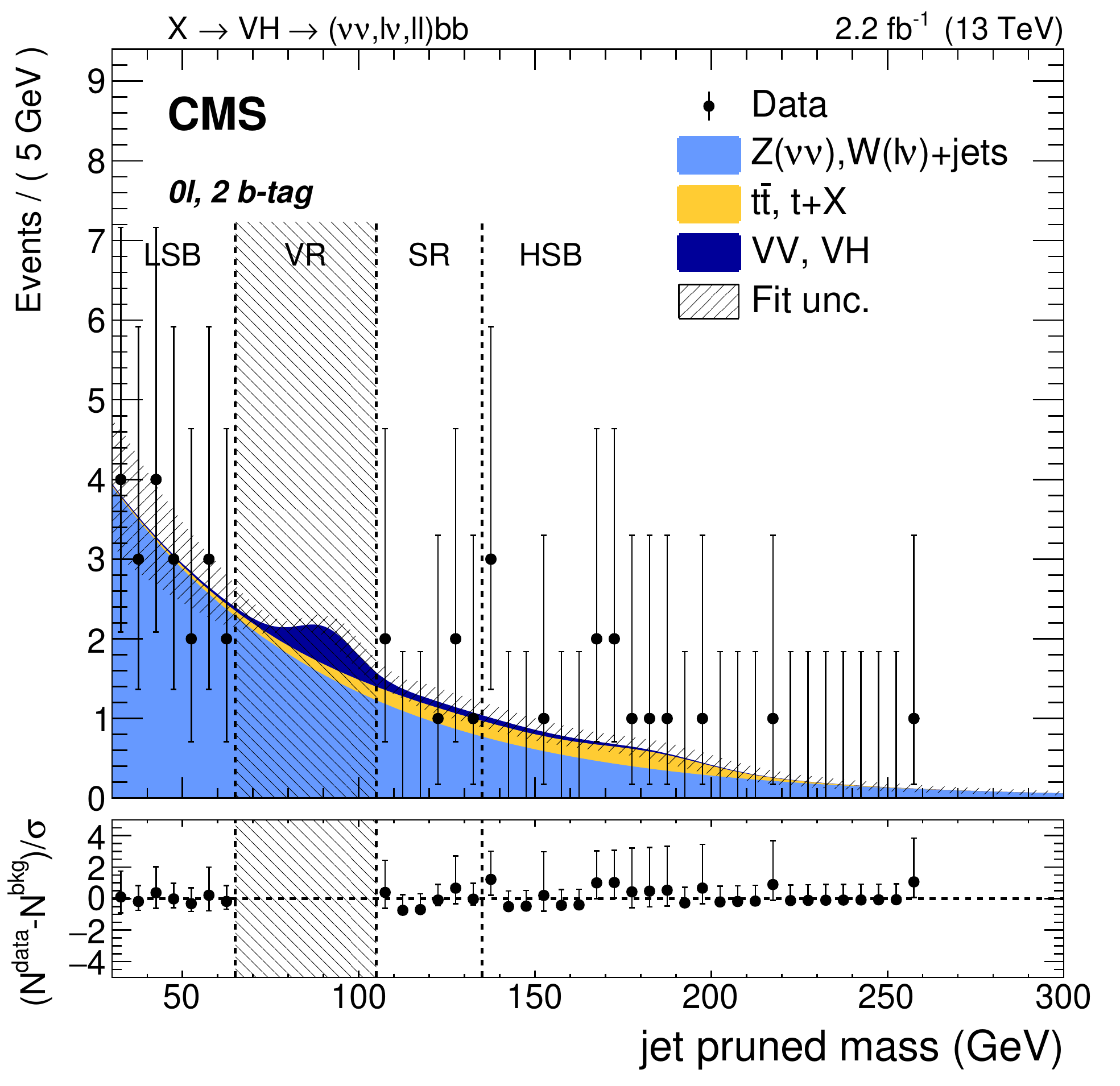}

\includegraphics[width=0.40\textwidth]{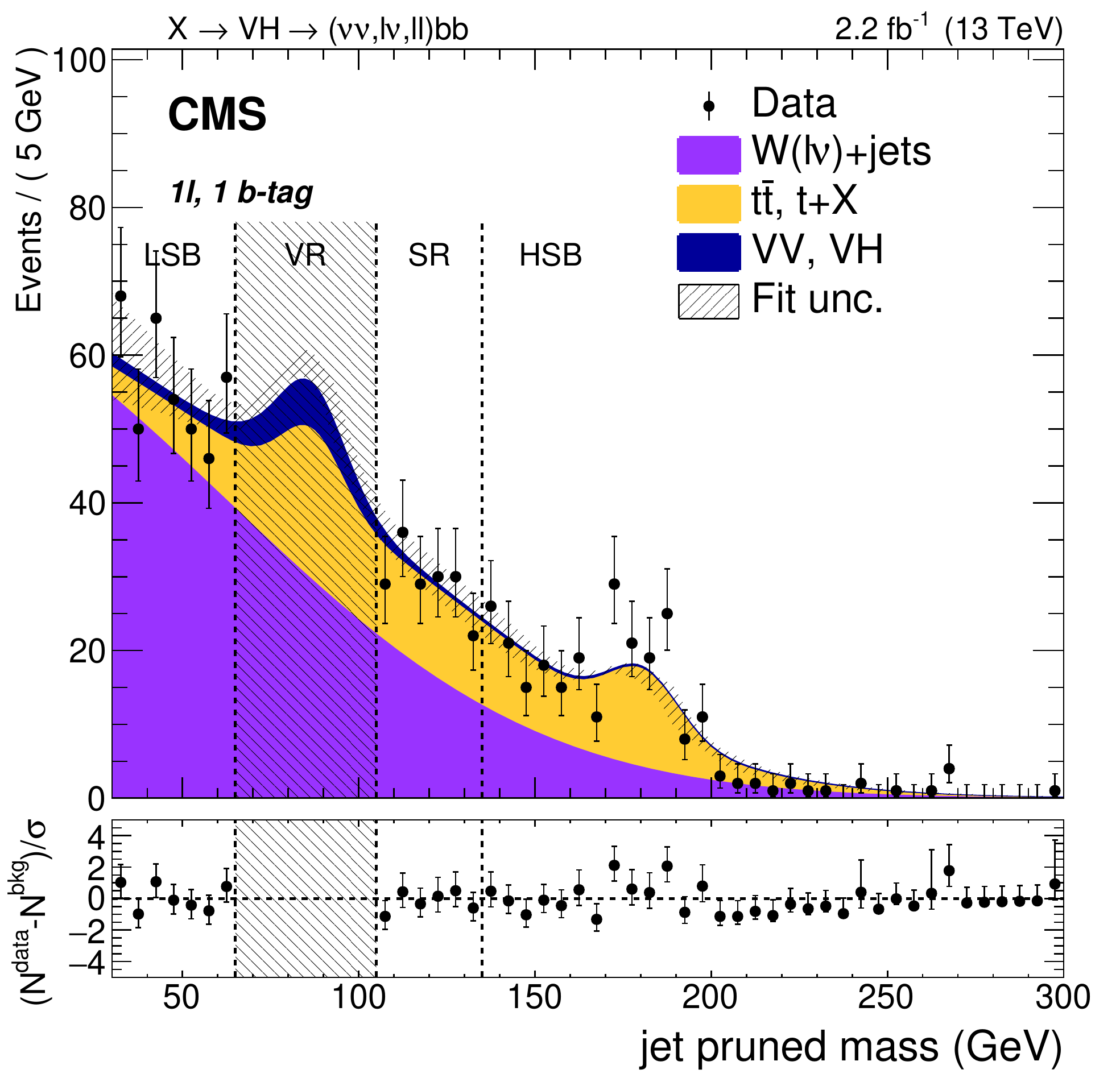}
\includegraphics[width=0.40\textwidth]{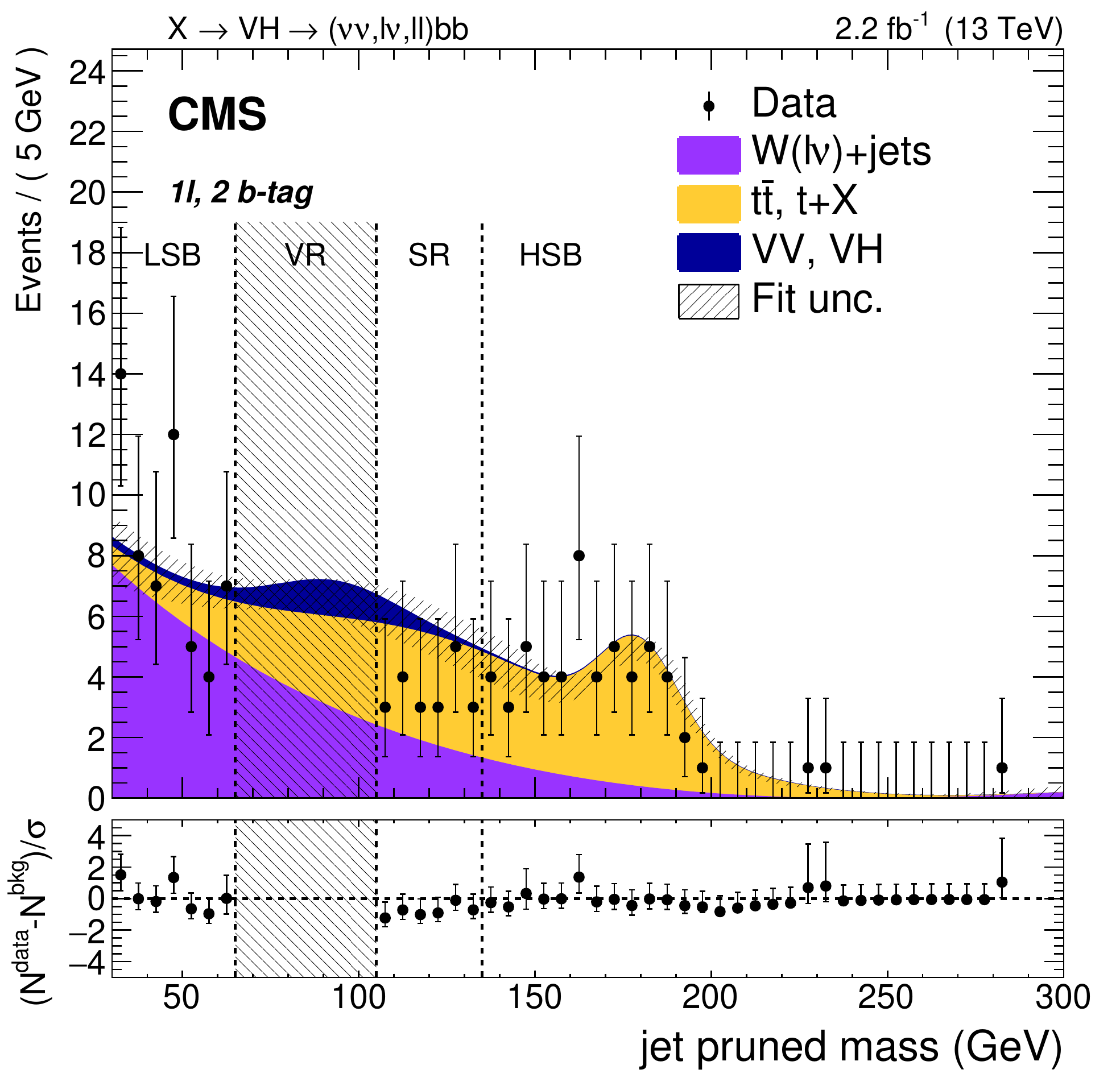}

\includegraphics[width=0.40\textwidth]{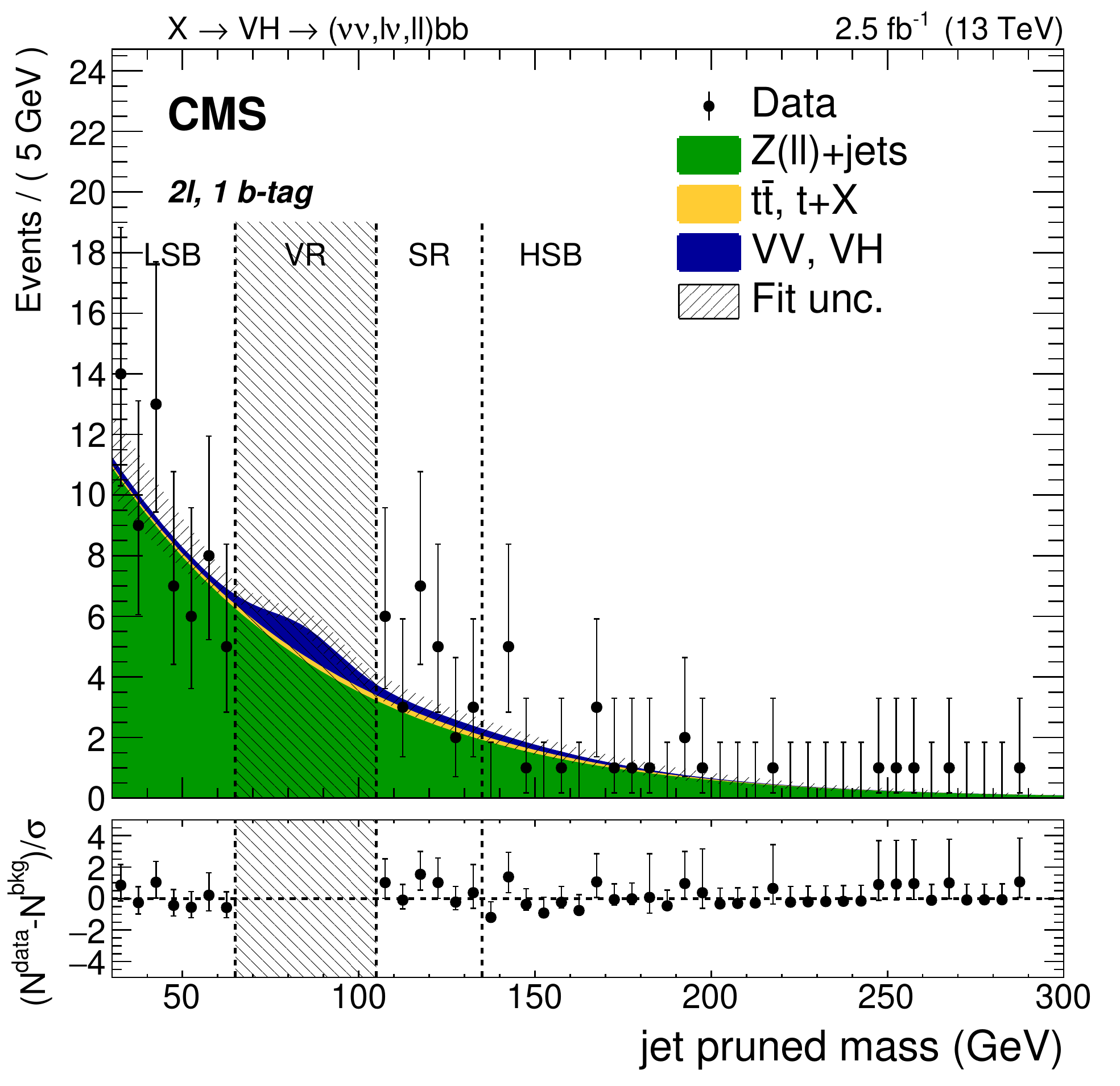}
\includegraphics[width=0.40\textwidth]{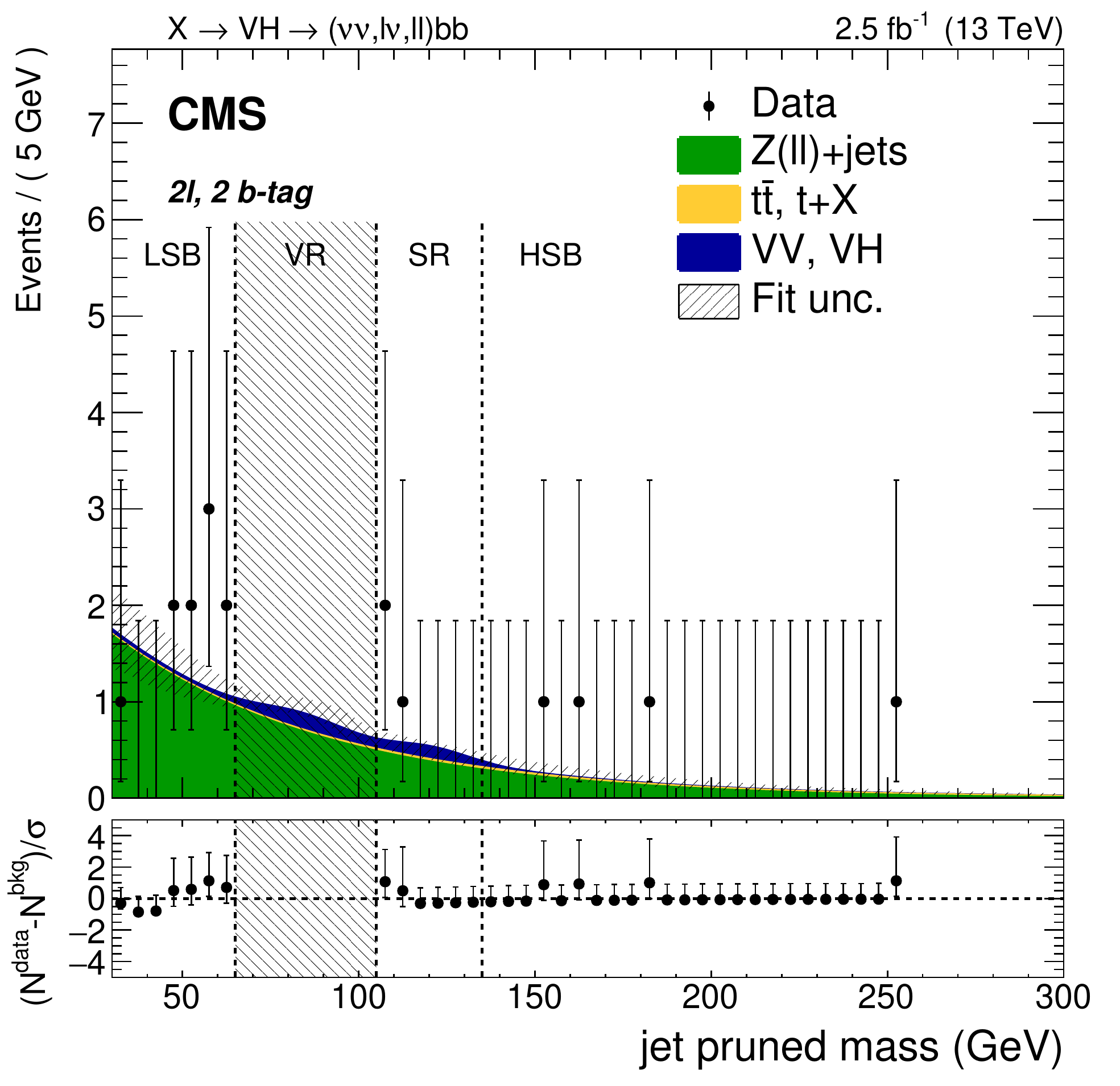}

  \caption{
    Pruned jet mass distribution of the leading AK8 jet in the $0\ell$ (upper), $1\ell$ (middle), and $2\ell$ (lower) categories, and separately for the 1 (\cmsLeft) and 2 (\cmsRight) b-tagged subjet selections. The shaded band represents the uncertainty from the fit to data in the pruned jet mass sidebands. The observed data are indicated by black markers. The dashed vertical lines separate the lower (LSB) and upper (HSB) sidebands, the \PW and \Z bosons mass region (VR), and the signal region (SR). The bottom panels report the pulls in each bin, $(N^\text{data}-N^\text{bkg})/\sigma$, where $\sigma$ is the Poisson uncertainty in data. The error bars represent the normalized Poisson errors on the data.
    \label{fig:mJ} }
\end{figure*}

The shape and normalization of the \mj distribution for the main \Vjets background is extracted from a fit of the sum of all contributing processes to the SB data, after fixing the shape and normalization of the subdominant backgrounds. The fits to the \mj distributions are shown in Fig.~\ref{fig:mJ}.
The normalization of the diboson processes is derived from simulation, while the top quark normalization is taken from the control regions with the exception of the dilepton channels.
The procedure is repeated selecting an alternative function to model the \mj distribution for the main background. The difference between the results obtained with the main and the alternative function is considered as a systematic uncertainty.
The number of expected and observed events in the SR are reported separately for each category in Table~\ref{tab:BkgNorm}.
A deficit of $2.4$ standard deviations is observed in the $1\mu$, 2 b tag category.

\begin{table*}[!htb]
  \topcaption{Expected and observed numbers of events in the signal region, for all event categories. Three separate sources of uncertainty in the expected numbers are reported: statistical uncertainty from the fit procedure (fit), the shape of the top quark and diboson background distributions (\ttbar, \VV), and the difference between the nominal and alternative function choice for the fit (alt. function).}\label{tab:BkgNorm}
  \centering
  \newcolumntype{x}{D{.}{.}{3.0}}
   \newcolumntype{y}{D{.}{.}{3.1}}
  \newcolumntype{z}{D{.}{.}{2.1}}
    \begin{tabular}{cc x y {c}@{\hspace*{5pt}} yzz}
      \multicolumn{2}{c}{\multirow{2}{*}{category}} & \multicolumn{2}{c}{events} && \multicolumn{3}{c}{uncertainties} \\
      \cline{3-4}\cline{6-8}
      & & \multicolumn{1}{c}{observed} & \multicolumn{1}{c}{expected} && \multicolumn{1}{c}{fit} & \multicolumn{1}{c}{\ttbar, \VV} & \multicolumn{1}{c}{alt. function} \\
      \hline
      \multirow{5}{*}{1 b tag} & $0\ell$ &  47  &  49.5 & &  \pm8.5  &  \pm0.4  &  \pm6.9  \\
      & $1\Pe$ & 57 & 73 && \pm23 & \pm1 & \pm6 \\
      & $1\mu$ & 119 & 123 && \pm8 & \pm1 & \pm5 \\
      & $2\Pe$ & 7 & 4.8 && \pm1.1 & \pm0.1 & \pm1.0 \\
      & $2\mu$ & 19 & 13.2 && \pm1.8 & \pm0.1 & \pm0.8 \\
      \hline
      \multirow{5}{*}{2 b tag} & $0\ell$ &  6  &  8.0  &&  \pm1.3  &  \pm0.2  &  \pm1.2  \\
      & $1\Pe$ &  7  &  8.7  &&  \pm1.0  &  \pm0.3  &  \pm0.5  \\
      & $1\mu$ &  14  &  29.5  &&  \pm3.4  &  \pm1.0  &  \pm0.9  \\
      & $2\Pe$ &  2  &  1.1  &&  \pm0.5  &  \pm0.1  &  \pm0.1  \\
      & $2\mu$ &  1  &  1.9  &&  \pm0.7  &  {<}0.1 &  \pm0.3  \\
      \hline
    \end{tabular}
\end{table*}

\begin{figure*}[!hbtp]\centering
  \includegraphics[width=0.40\textwidth]{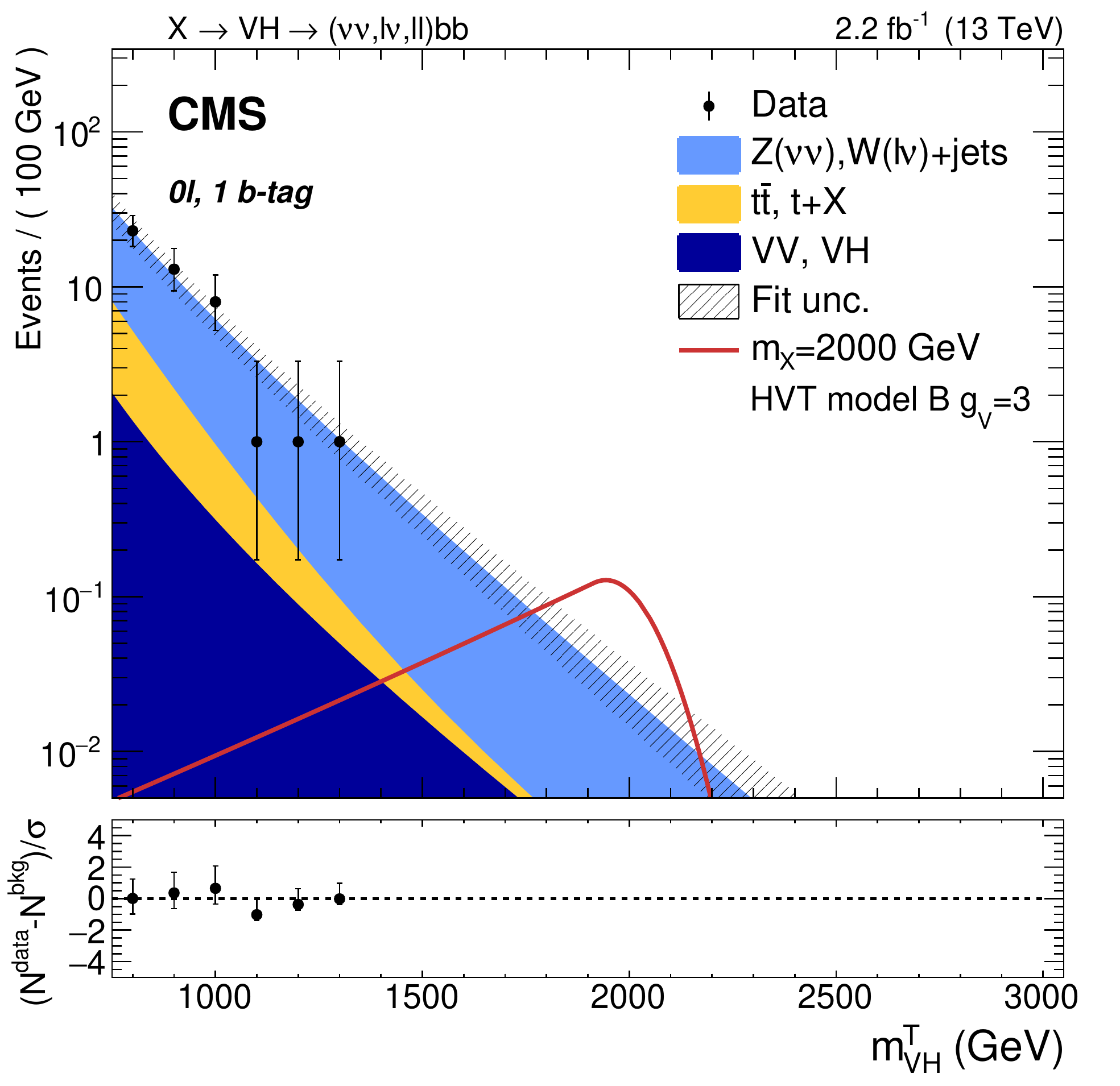}
  \includegraphics[width=0.40\textwidth]{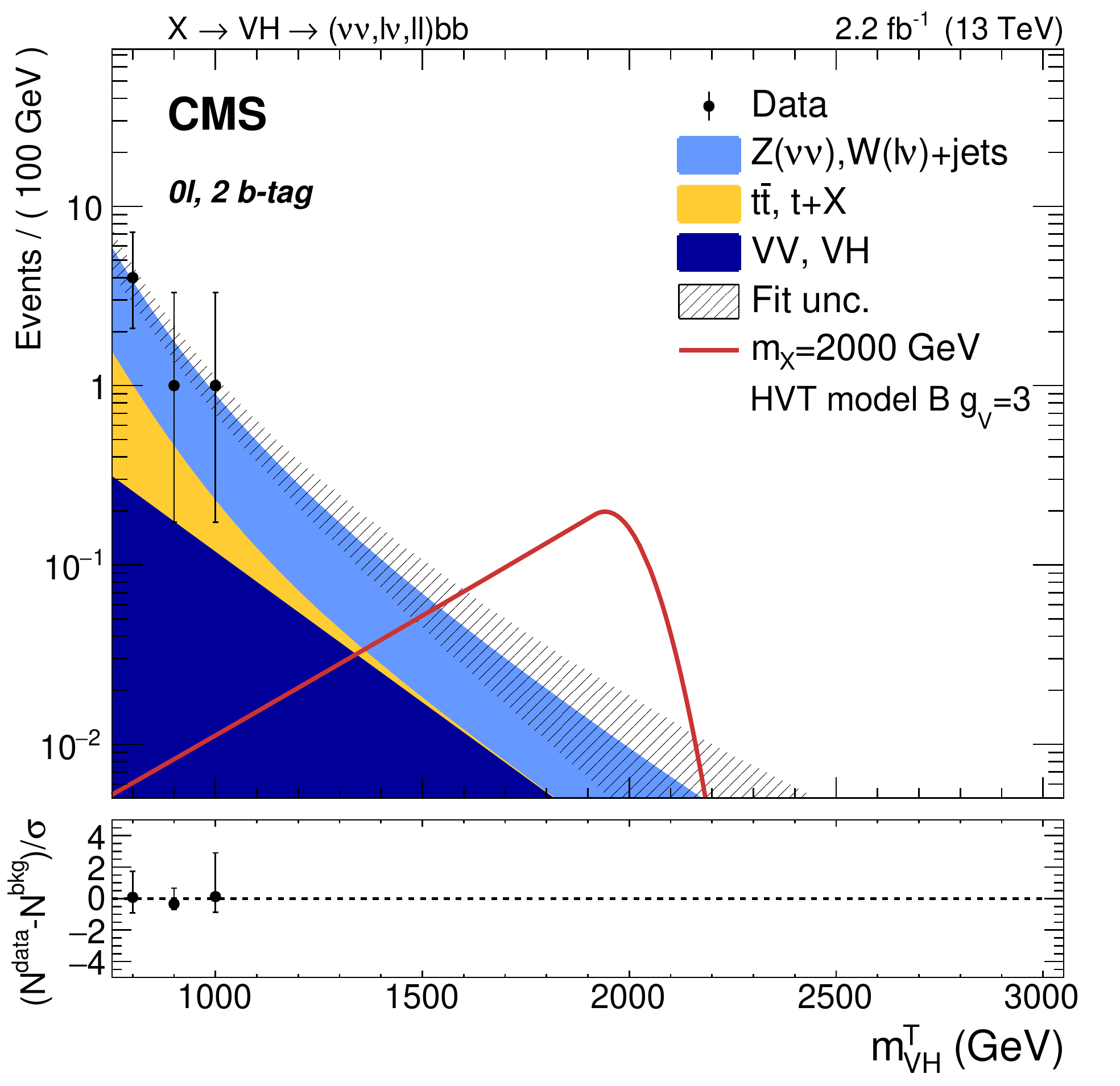}

  \includegraphics[width=0.40\textwidth]{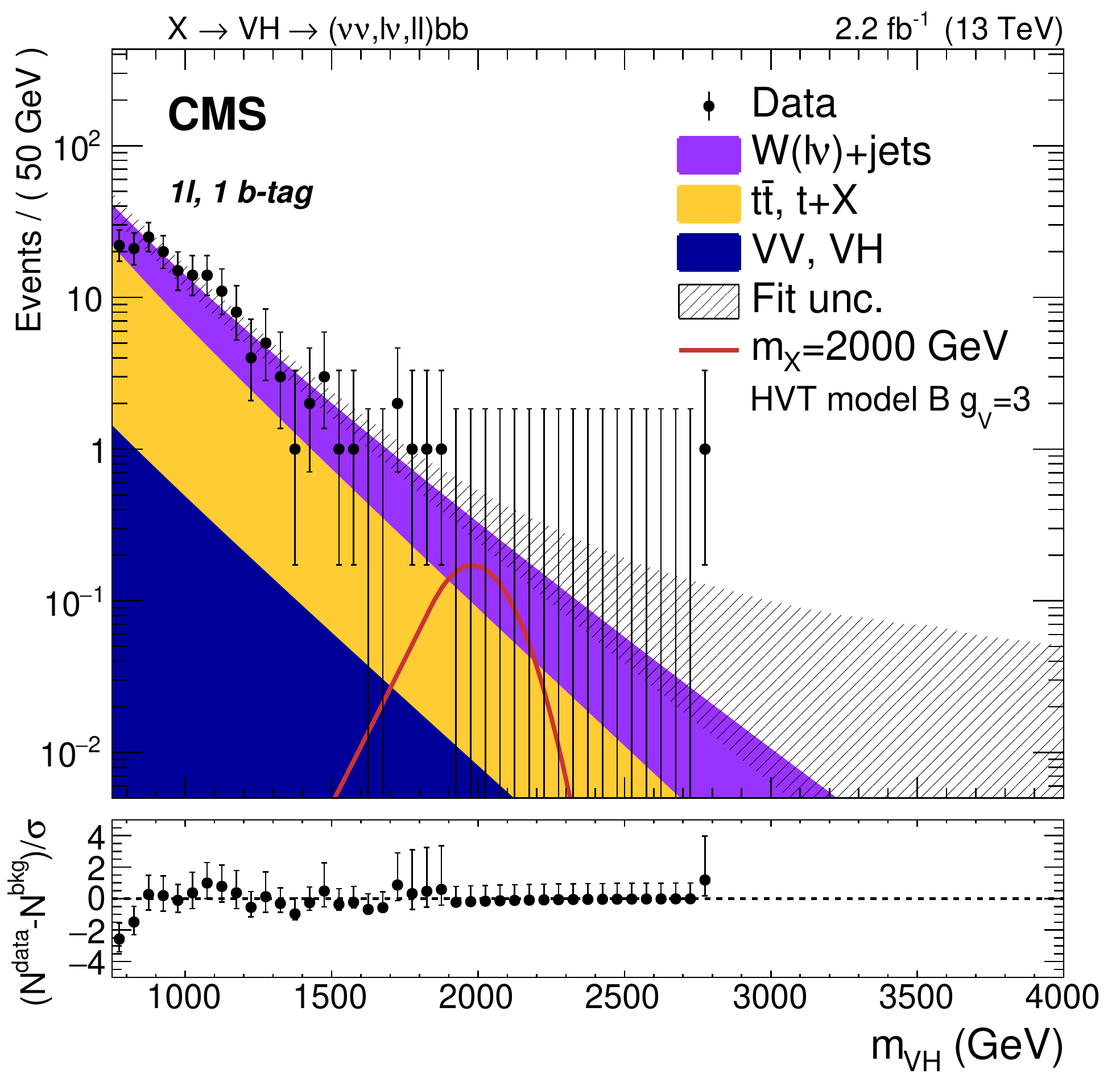}
  \includegraphics[width=0.40\textwidth]{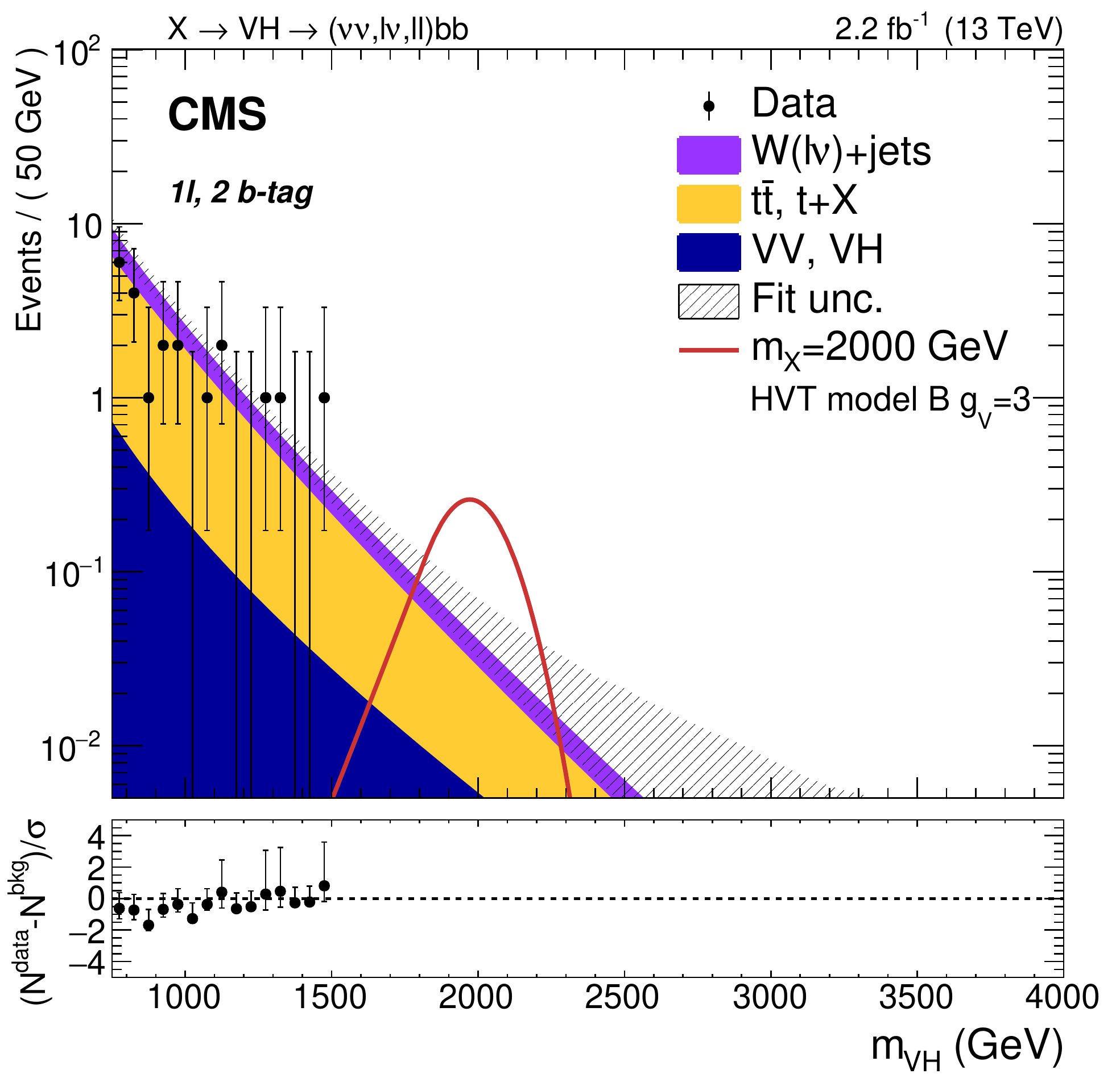}

  \includegraphics[width=0.40\textwidth]{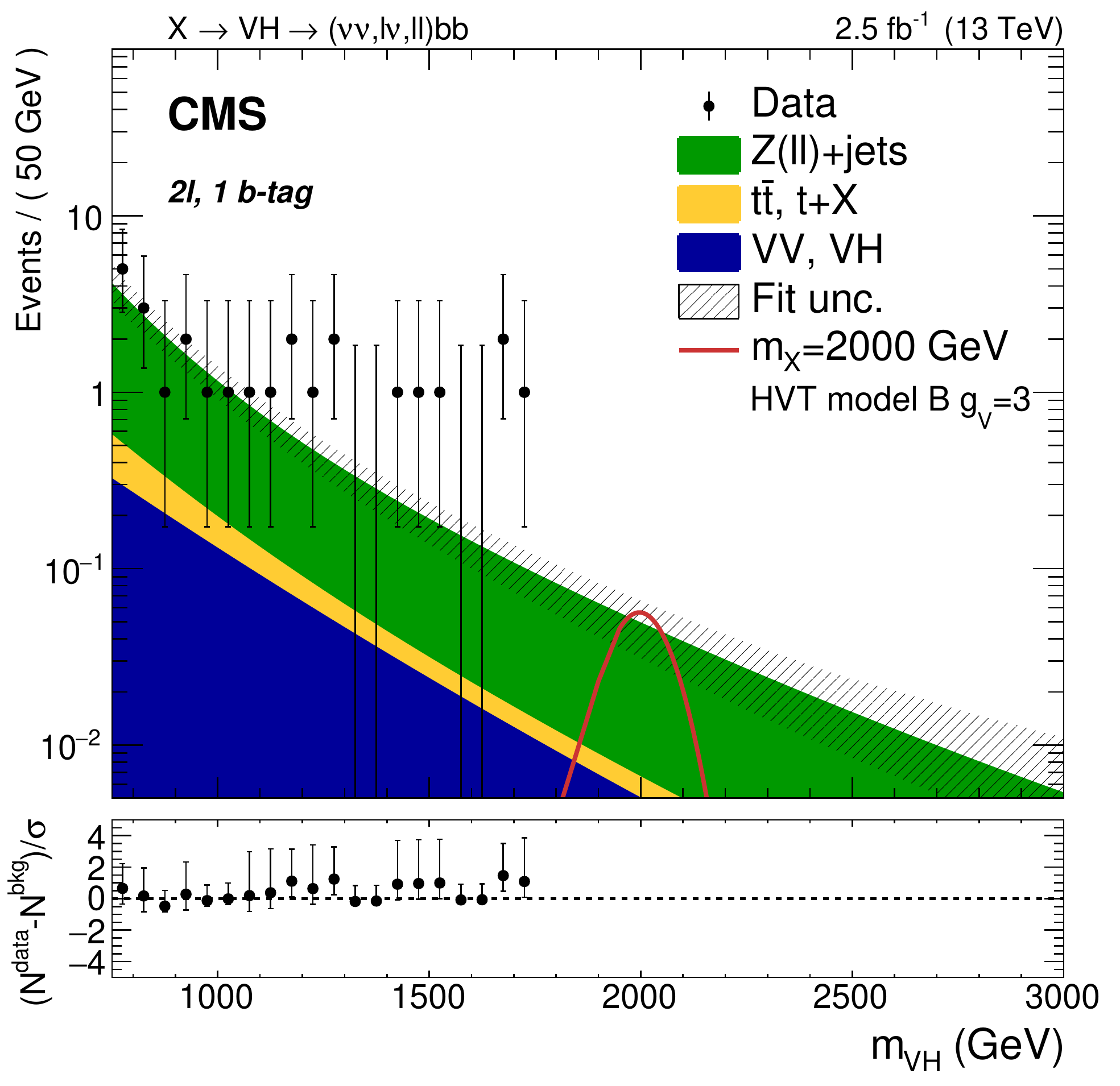}
  \includegraphics[width=0.40\textwidth]{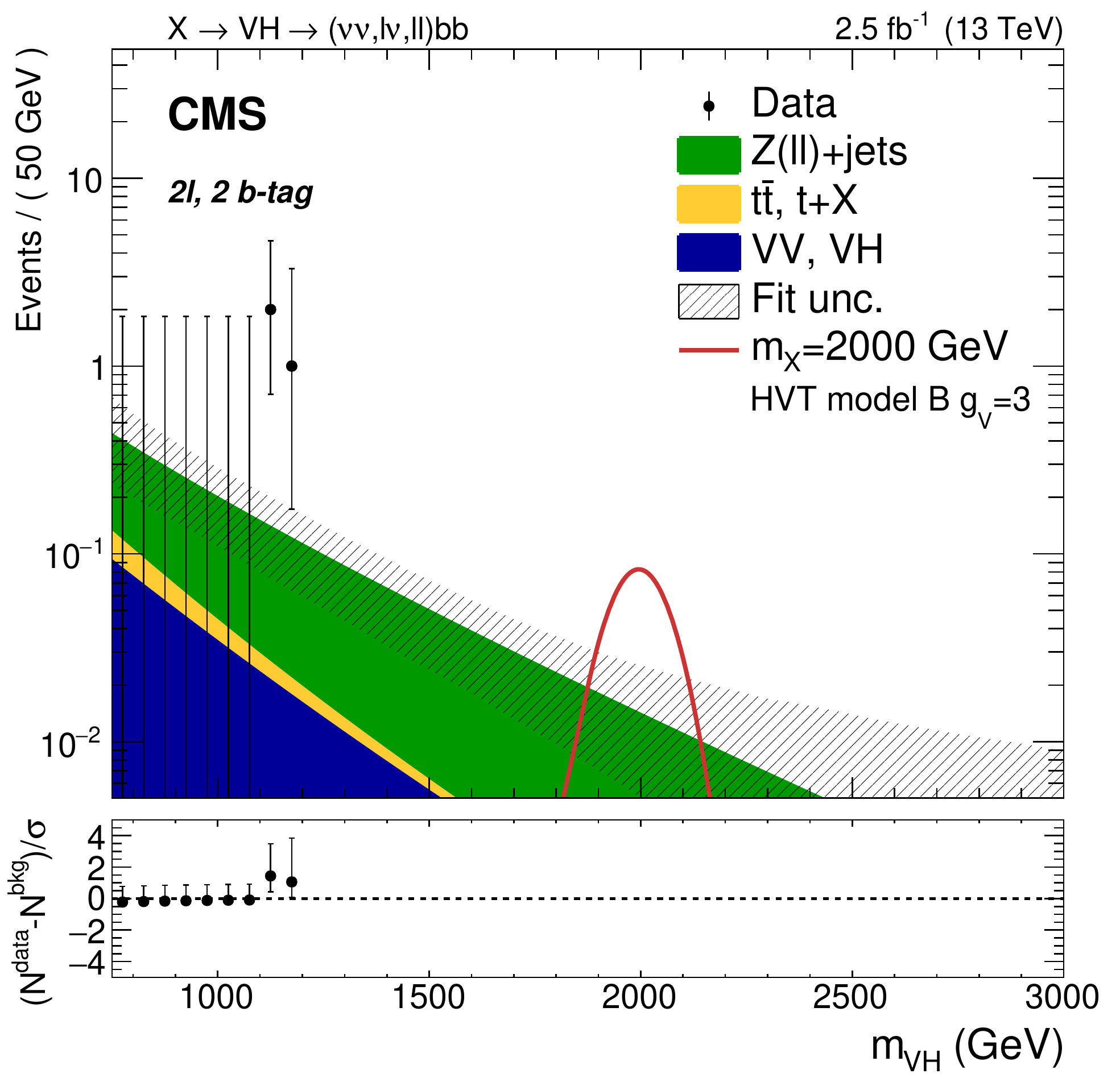}

  \caption{
    Resonance candidate mass \mVH distributions in the $0\ell$ (upper), $1\ell$ (middle), and $2\ell$ (lower) categories, and separately for the 1 (\cmsLeft) and 2 (\cmsRight) b-tagged subjet selections. The expected background events are shown with the filled area, and the shaded band represents the total background uncertainty. The observed data are indicated by black markers, and the potential contribution of a resonance with \mX = 2000\GeV produced in the context of the HVT model B with $\gV=3$ is shown with a solid red line.
    The bottom panels report the pulls in each bin, $(N^\text{data}-N^\text{bkg})/\sigma$, where $\sigma$ is the Poisson uncertainty in data. The error bars represent the normalized Poisson errors on the data.
    \label{fig:mX} }
\end{figure*}

The shape of the \Vjets background distribution in the \mVH variable is obtained via a transfer function determined from simulation as:
\begin{equation}
\alpha(\mVH) = \frac{N_\mathrm{SR}^{\text{sim},\Vjets}(\mVH)}{N_\mathrm{SB}^{\text{sim},\Vjets}(\mVH)}
\end{equation}
where $N_\mathrm{SR}^{\text{sim},\Vjets}(\mVH)$, $N_\mathrm{SB}^{\text{sim},\Vjets}(\mVH)$ are two-parameter probability density functions determined from the \mVH spectra in the SR and the SB of the simulated \Vjets sample, respectively. The ratio $\alpha(\mVH)$ accounts for the correlations and the small kinematic differences involved in the interpolation from the sidebands to the SR, and is largely independent of the shape uncertainties and the assumptions on the overall cross section.
The shape of the main background is extracted from data in the \mj sidebands, after multiplying the obtained distribution by the $\alpha(\mVH)$ ratio. The overall predicted background distribution in the SR, $N_\mathrm{SR}^\text{pred}(\mVH)$, is given by the following relation:
\ifthenelse{\boolean{cms@external}}{
\begin{multline}
N_\mathrm{SR}^\text{pred}(\mVH) =  N_\mathrm{SB}^\text{obs,\Vjets}(\mVH) \, \alpha(\mVH) \\+ N_\mathrm{SR}^\text{sim,\ttbar}(\mVH) + N_\mathrm{SR}^\text{sim,\VV}(\mVH)
\end{multline}
}{
\begin{equation}
N_\mathrm{SR}^\text{pred}(\mVH) =  N_\mathrm{SB}^\text{obs,\Vjets}(\mVH) \, \alpha(\mVH) + N_\mathrm{SR}^\text{sim,\ttbar}(\mVH) + N_\mathrm{SR}^\text{sim,\VV}(\mVH)
\end{equation}
}
where $N_\mathrm{SB}^\text{obs,\Vjets}(\mVH)$ is the probability distribution function obtained from a fit to data in the \mj sidebands of the sum of the background components, and $N_\mathrm{SR}^\text{sim,\ttbar}(\mVH)$ and $N_\mathrm{SR}^\text{sim,\VV}(\mVH)$ are the \ttbar and diboson components, respectively, fixed to the shapes and normalizations derived from the simulated samples and control regions.
The observed data in the SR are in agreement with the predicted background, as shown in Fig.~\ref{fig:mX}.

The validity and robustness of this method is tested on data by splitting the lower \mj sideband in two and predicting shape and normalization of the intermediate sideband from the lower and upper sidebands. The number of events and distributions found in data are compatible with the prediction within the systematic uncertainties.

The shape of the reconstructed signal mass distribution is extracted from the simulated signal samples. The signal shape is parametrized separately for each channel with a Gaussian peak and a power law to model the lower tails.
The resolution of the reconstructed \mVH is given by the width of the Gaussian core for the $1\ell$ and $2\ell$ channels and by the RMS of the \mtVH distribution in the $0\ell$ channel, and is found to be 10--16\%, 8--5\%, 5--3\% of \mX in the $0\ell$, $1\ell$, and $2\ell$ channels, respectively, when going from low to high resonance masses.

\section{Systematic uncertainties}
\label{sec:sys}

The sensitivity of this analysis is limited by statistical rather than systematic uncertainties.

The systematic uncertainty in the \Vjets background yield is dominated by the statistical uncertainty associated with the number of data events in the \mj sideband. Minor contributions arise from the propagation of the uncertainties in the shape of the function modeling the \mj distributions of the \ttbar and \VV backgrounds. The \ttbar and \ST normalization uncertainty, in the $0\ell$ and $1\ell$ categories, originates from the limited number of events in the top quark control regions. The diboson normalization uncertainty depends on the propagation of the theoretical uncertainties in the relevant phase space, and is estimated to be 20\%. Given the rather large scale factor observed in the $0\ell$, 2 b tag \ttbar control region, the top quark normalization uncertainty in the $2\ell$ category is conservatively taken to be 50\%.

The uncertainties in the \Vjets background shape are estimated from the covariance matrix of the fit to data of the \mVH distribution in the sideband regions and from the uncertainties in the modeling of the $\alpha(\mVH)$ ratio, which depends on the number of data and simulation events, respectively.

Other sources of uncertainty affect both the normalization and shape of the simulated signal and the subdominant backgrounds.
The uncertainties in the trigger efficiency and the electron, muon, and $\tau$ lepton reconstruction, identification, and isolation are evaluated through specific studies of events with dilepton masses in the region of the \Z peak, and amount to a 6--8\% uncertainty for the categories with charged leptons, and 3\% in the $0\ell$ categories. In the $1\ell$ and $2\ell$ categories, the lepton energy scale and resolution are propagated to the signal shape, and the resulting uncertainties in the mean and the width of the signal model are estimated to be as large as 16\% and 10\%, respectively, depending on the lepton flavor and signal mass.
The jet energy scale and resolution~\cite{Khachatryan:2016kdb} affect both shape and selection efficiencies. The jet energy corrections, propagated to the jet mass, are also taken into account, and are responsible for a 5\% variation in the background, and a variation of 1--3\%, depending on the mass hypothesis, in the number of signal events. The jet energy resolution accounts for an additional 2--3\% uncertainty. The effects are propagated to the \mVH distributions and considered as uncertainties in the subdominant backgrounds and signal samples. As a result, in the signal sample a 0.3\% uncertainty is assigned to the mean of the signal shape, and 1.0\% to the width.

The efficiency for signal events to enter the SR jet mass window is evaluated with \HERWIG~\cite{Bellm:2015jjp,Bahr:2008pv} as an alternative showering algorithm. The 7\% difference observed with respect to the default \PYTHIA showering is taken to be the systematic uncertainty.

Uncertainties on the b tagging efficiency~\cite{CMS-PAS-BTV-15-001} represent the largest source of normalization uncertainty for samples that are not normalized to data. For the signal efficiency, these uncertainties in the yield of between 4--15\% and 8--30\%, depending on \mVH, are estimated in the 1 and 2 b-tagged subjet categories, respectively; for background events, respective uncertainties of 5 and  12\% are found in the two cases. An additional 10\% b tagging uncertainty is assigned to the \ttbar background to account for the extrapolation from the top quark control region to the SR.

The factorization and renormalization scale uncertainties associated with the event generators are estimated by varying the corresponding scales up and down by a factor of 2, and are responsible for a 5\% normalization variation in the estimated diboson background. The effect of these scale uncertainties is propagated to the \ttbar and \VV background distributions, and the difference in the \mVH distribution parameters is taken as an additional shape uncertainty. The effect on the signal shape modeling is negligible, and the resulting normalization uncertainty is 4--12\%, depending on \mVH.

Additional systematic uncertainties affecting the normalization of backgrounds and signal from pileup contributions (3 and 0.5\%), integrated luminosity (2.7\%)~\cite{CMS:2016eto}, \MET scale and resolution (1\% in the $0\ell$ channel), and the choice of PDFs~\cite{Butterworth:2015oua} (3\% for acceptance, and 4--18\% for signal normalization) are also included in the analysis.

\section{Results and interpretation}
\label{sec:res}

\begin{figure}[!htb]\centering
    \includegraphics[width=\cmsFigWidth]{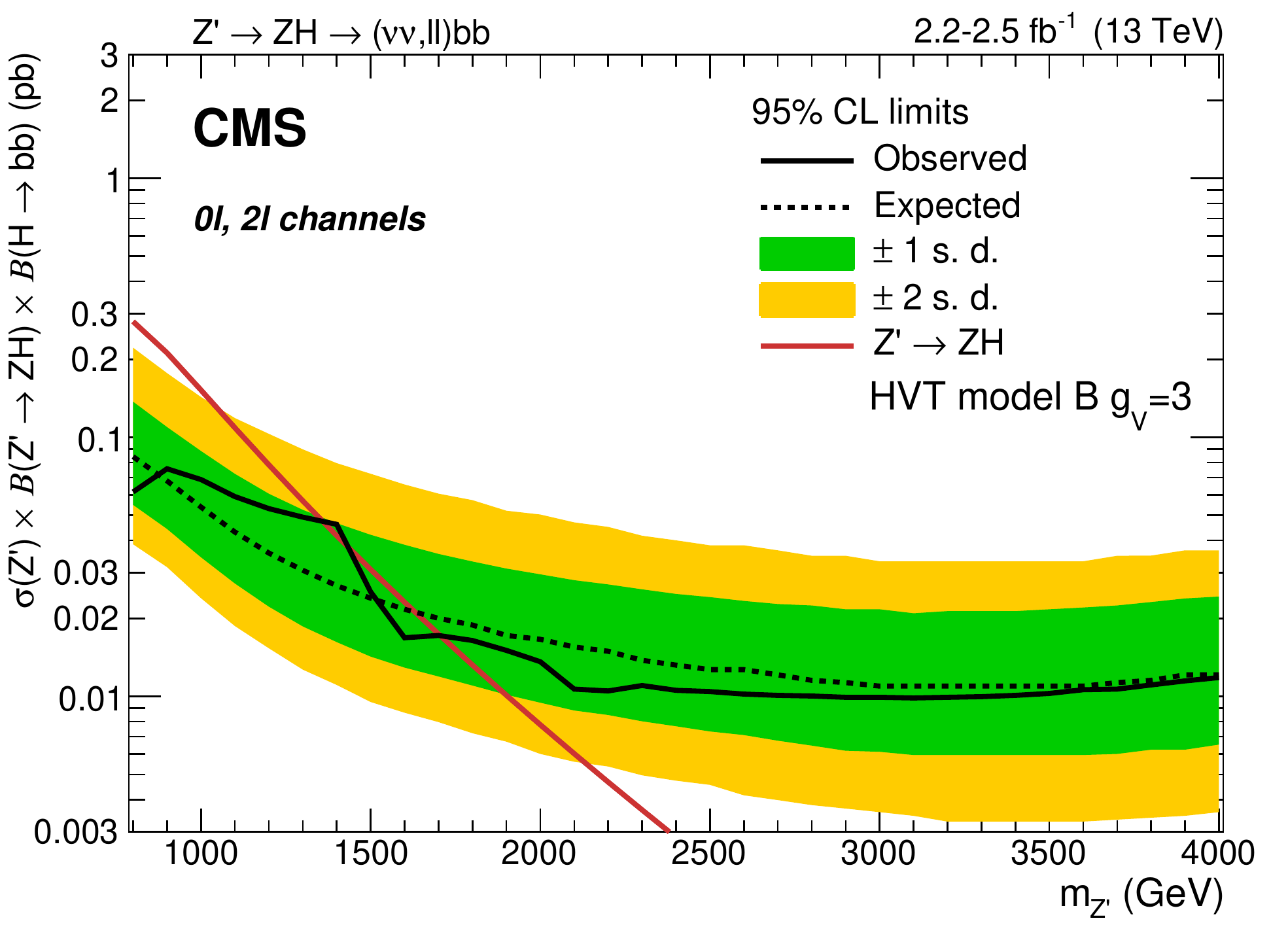}
    \includegraphics[width=\cmsFigWidth]{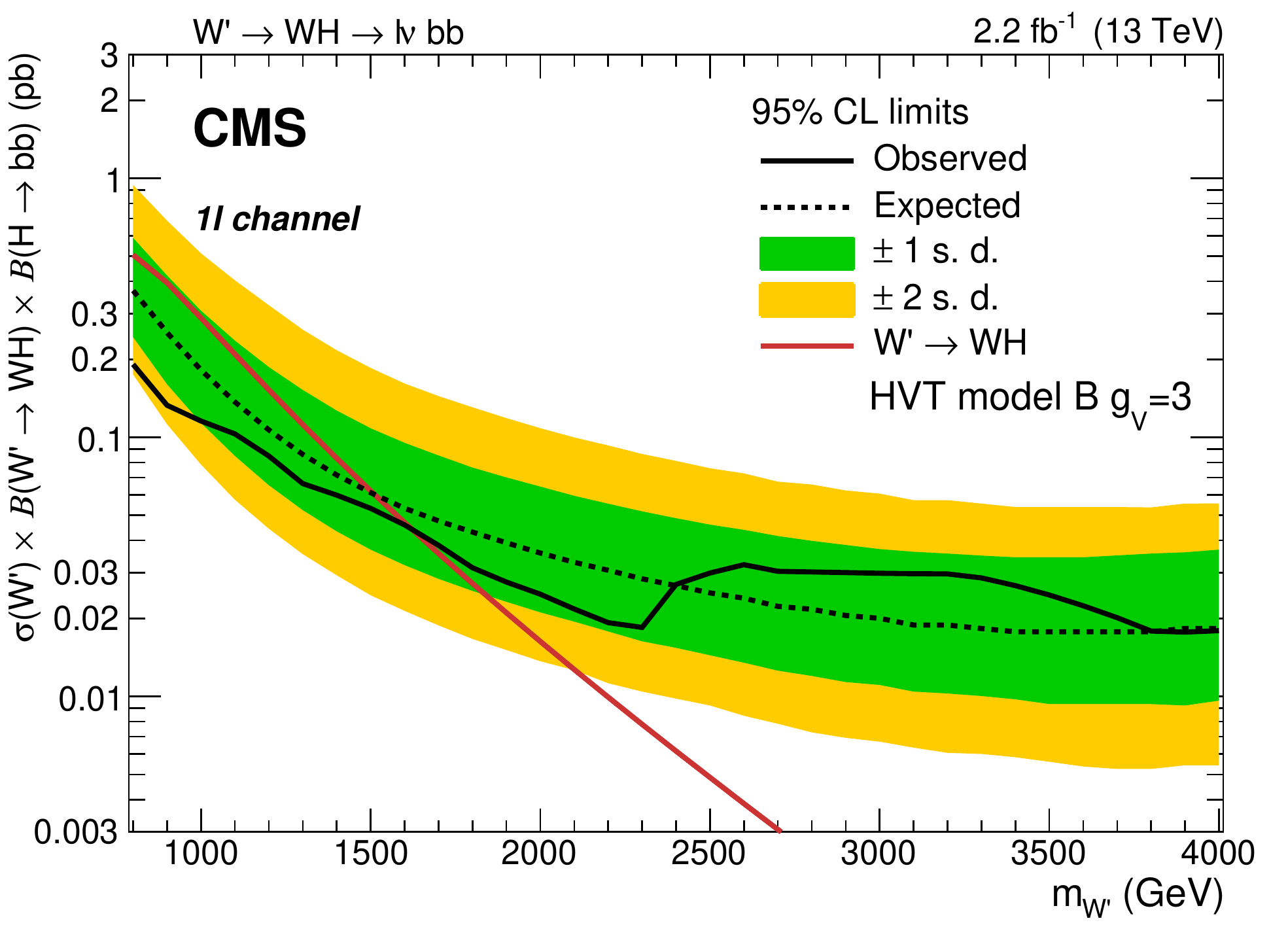}
    \caption{Observed and expected 95\% CL upper limits on $\sigma(\PZpr) \, \B(\PZpr\to\Z\h) \, \B(\htobb)$ (\cmsLeft) and $\sigma(\PW') \, \B(\PW'\to\PW\h) \, \B(\htobb)$ (\cmsRight) as a function of the resonance mass for a single narrow spin-1 resonance, including all statistical and systematic uncertainties. The inner green and outer yellow bands represent the ${\pm}1$ and ${\pm}2$ standard deviation uncertainties on the expected limit. The red solid curve corresponds to the cross sections predicted by the HVT model~B with $\gV=3$.
    }
  \label{fig:limit}
\end{figure}

Results are obtained from a combined signal and background fit to the unbinned \mVH distribution, based on a profile likelihood. Systematic uncertainties are treated as nuisance parameters and are profiled in the statistical interpretation~\cite{CLS1,CLS2,CMS-NOTE-2011-005,Asymptotic}. The background-only hypothesis is tested against the $\X\to\VH$ signal in the ten categories.
The asymptotic modified frequentist method is used to determine limits at 95\% confidence level (CL) on the contribution from signal.
Limits are derived on the product of the cross section for a heavy vector boson \X and the branching fractions for the decays $\X\to\VH$ and $\htobb$, denoted $\sigma(\X) \, \B(\X\to\VH) \, \B(\htobb)$. No specific assumption is made on $\B(\htobb)$, since this decay channel has not yet been measured.
The $0\ell$ and $2\ell$ categories are combined to provide upper limits for the case where \X is a heavy spin-1 vector singlet $\PZpr$, in the narrow-width approximation. Similarly the $1\ell$ categories are combined to provide limits for the case where \X is a heavy $\PW'$. The exclusion limits are reported in Fig.~\ref{fig:limit}. These limits are verified with the modified frequentist $\mathrm{CL}_\mathrm{s}$ method, obtaining results compatible with those obtained with the asymptotic formula.

\begin{figure}[!htb]\centering
    \includegraphics[width=\cmsFigWidth]{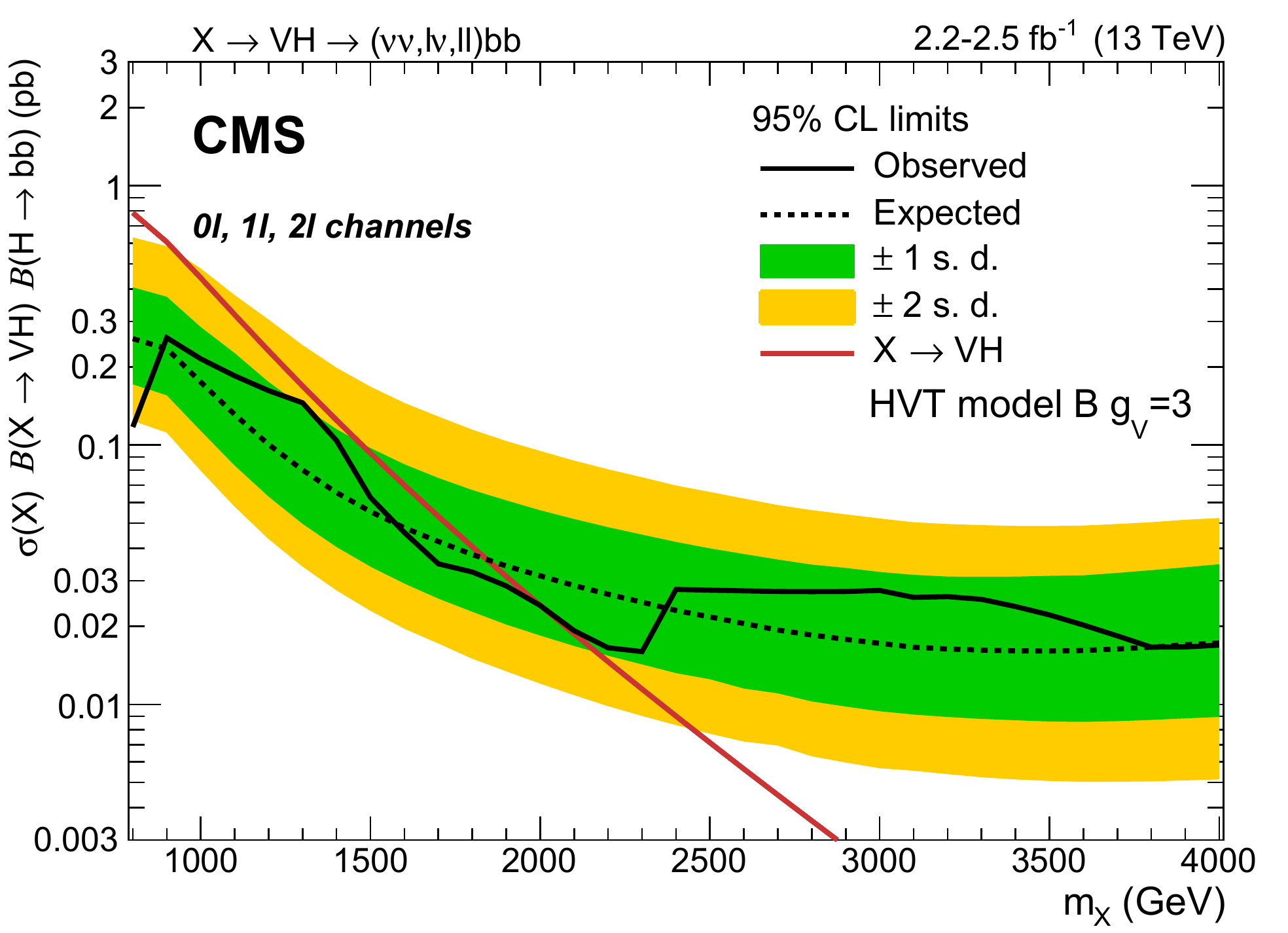}
    \caption{Observed and expected 95\% CL upper limit with the ${\pm}1$ and ${\pm}2$ standard deviation uncertainty bands on $\sigma(\X) \, \B(\X\to\VH) \, \B(\htobb)$  in the HVT model~B benchmark scenario with $\gV=3$ as a function of the resonance mass, for the combination of all the considered channels.}
  \label{fig:xvh}
\end{figure}

The result of this study is primarily interpreted in the context of a simplified model with a triplet of heavy vector bosons ($\V^\pm$, $\V^0$)~\cite{Pappadopulo2014}.
The predictions of the benchmark model~B are superimposed on the exclusion limits in Fig.~\ref{fig:limit}.
All the $0\ell$, $1\ell$, and $2\ell$ channels are combined to put stringent exclusion limits on the HVT model, scenario~B, assuming the $\PZpr$ and $\PW'$ cross sections as predicted by the model. There are normalization increases caused by event migration between the leptonic channels, which are estimated to be 5--10\% in the $0\ell$ channel, due to mis-assigned $\PW'$ events, and less than 1\%  in the $1\ell$ channel, due to mis-assigned $\PZpr$ events. Figure~\ref{fig:xvh} presents the exclusion limits as a function of the heavy triplet mass. A resonance with $\mX \lesssim 2.0\TeV$ is excluded at 95\% CL in the HVT model~B.

\begin{figure}[!htb]\centering
    \includegraphics[width=\cmsFigWidth]{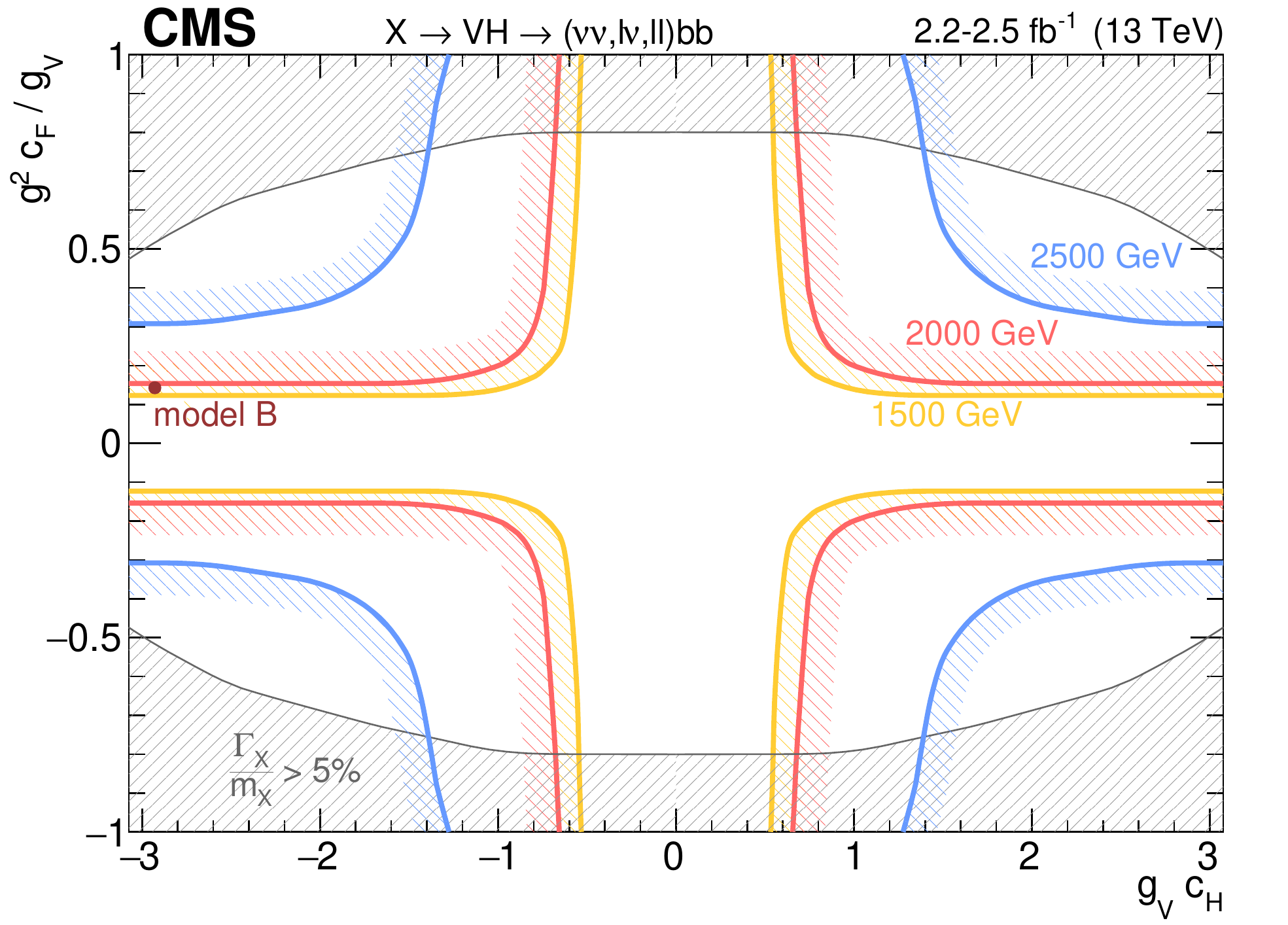}
    \caption{Observed exclusion in the HVT parameter plane $\left[ \gV \cH, \ g^2 \cF / \gV \right]$ for three different resonance masses (1.5, 2.0, and 2.5\TeV). The parameter $\gV$ represents the coupling strength of the new interaction, $\cH$ the coupling between the HVT bosons and the Higgs boson and longitudinally polarized SM vector bosons, and $\cF$ the coupling between the heavy vector bosons and the SM fermions. The benchmark scenario~B with $\gV=3$ is represented by the red point. The gray shaded area corresponds to the region where the resonance natural width is predicted to be larger than the typical experimental resolution (5\%), and thus the narrow-width approximation breaks down.}
  \label{fig:hvt}
\end{figure}

The exclusion limit shown in Fig.~\ref{fig:xvh} can be interpreted as a limit in the $\left[ \gV \cH, \ g^2 \cF / \gV \right]$ plane of the HVT parameters, where $g$ represents the electroweak coupling constant. The excluded region of the parameter space for narrow resonances relative to the combination of all the considered channels is shown in Fig.~\ref{fig:hvt}. The fraction of the parameter space where the natural width of the resonances is larger than the typical experimental resolution of 5\%, and thus the narrow width approximation is not valid, is also indicated in Fig.~\ref{fig:hvt}.
The exclusion of the parameter space significantly improves on the reach of $\sqrt{s}=8\TeV$ searches in the $1\ell$~\cite{Khachatryan:2016yji} and all-hadronic channels~\cite{Khachatryan:2015bma}. The sensitivity is equivalent within the statistical and systematic uncertainties to the corresponding $\sqrt{s}=13\TeV$ search from ATLAS~\cite{Aaboud:2016lwx}.

\section{Summary}
\label{sec:conclusions}

A search for a heavy resonance with mass between 800 and 4000\GeV, decaying into a vector boson and a Higgs boson, has been described. The data samples were collected by the CMS experiment at $\sqrt{s}=13\TeV$ during 2015, and correspond to integrated luminosities of 2.2--2.5 \fbinv, depending on the channel. The final states explored include the leptonic decay modes of the vector boson, events with zero ($\Z \to \nu\nu$), exactly one ($\PW \to \ell\nu$), and two ($\Z \to \ell\ell$) charged leptons, with $\ell = \Pe,\mu$. Higgs bosons are reconstructed from their decays to \bbbar pairs. Depending on the resonance mass, upper limits in the range 10--200 \unit{fb} are set on the product of the cross section for a narrow spin-1 resonance and the branching fractions for the decay of the resonance into a Higgs and a vector boson, and for the decay of the Higgs boson into a pair of b quarks. Resonances with masses lower than 2\TeV are excluded within the heavy vector triplet model in the benchmark scenario~B with $\gV=3$. These results represent a significant reduction in the allowed parameter space for the large number of models generalized within the heavy vector triplet framework.

\begin{acknowledgments}
We congratulate our colleagues in the CERN accelerator departments for the excellent performance of the LHC and thank the technical and administrative staffs at CERN and at other CMS institutes for their contributions to the success of the CMS effort. In addition, we gratefully acknowledge the computing centers and personnel of the Worldwide LHC Computing Grid for delivering so effectively the computing infrastructure essential to our analyses. Finally, we acknowledge the enduring support for the construction and operation of the LHC and the CMS detector provided by the following funding agencies: BMWFW and FWF (Austria); FNRS and FWO (Belgium); CNPq, CAPES, FAPERJ, and FAPESP (Brazil); MES (Bulgaria); CERN; CAS, MoST, and NSFC (China); COLCIENCIAS (Colombia); MSES and CSF (Croatia); RPF (Cyprus); SENESCYT (Ecuador); MoER, ERC IUT, and ERDF (Estonia); Academy of Finland, MEC, and HIP (Finland); CEA and CNRS/IN2P3 (France); BMBF, DFG, and HGF (Germany); GSRT (Greece); OTKA and NIH (Hungary); DAE and DST (India); IPM (Iran); SFI (Ireland); INFN (Italy); MSIP and NRF (Republic of Korea); LAS (Lithuania); MOE and UM (Malaysia); BUAP, CINVESTAV, CONACYT, LNS, SEP, and UASLP-FAI (Mexico); MBIE (New Zealand); PAEC (Pakistan); MSHE and NSC (Poland); FCT (Portugal); JINR (Dubna); MON, RosAtom, RAS, and RFBR (Russia); MESTD (Serbia); SEIDI and CPAN (Spain); Swiss Funding Agencies (Switzerland); MST (Taipei); ThEPCenter, IPST, STAR, and NSTDA (Thailand); TUBITAK and TAEK (Turkey); NASU and SFFR (Ukraine); STFC (United Kingdom); DOE and NSF (USA).

\hyphenation{Rachada-pisek} Individuals have received support from the Marie-Curie program and the European Research Council and EPLANET (European Union); the Leventis Foundation; the A. P. Sloan Foundation; the Alexander von Humboldt Foundation; the Belgian Federal Science Policy Office; the Fonds pour la Formation \`a la Recherche dans l'Industrie et dans l'Agriculture (FRIA-Belgium); the Agentschap voor Innovatie door Wetenschap en Technologie (IWT-Belgium); the Ministry of Education, Youth and Sports (MEYS) of the Czech Republic; the Council of Science and Industrial Research, India; the HOMING PLUS program of the Foundation for Polish Science, cofinanced from European Union, Regional Development Fund, the Mobility Plus program of the Ministry of Science and Higher Education, the National Science Center (Poland), contracts Harmonia 2014/14/M/ST2/00428, Opus 2013/11/B/ST2/04202, 2014/13/B/ST2/02543 and 2014/15/B/ST2/03998, Sonata-bis 2012/07/E/ST2/01406; the Thalis and Aristeia programs cofinanced by EU-ESF and the Greek NSRF; the National Priorities Research Program by Qatar National Research Fund; the Programa Clar\'in-COFUND del Principado de Asturias; the Rachadapisek Sompot Fund for Postdoctoral Fellowship, Chulalongkorn University and the Chulalongkorn Academic into Its 2nd Century Project Advancement Project (Thailand); and the Welch Foundation, contract C-1845.
\end{acknowledgments}
\bibliography{auto_generated}

\cleardoublepage \appendix\section{The CMS Collaboration \label{app:collab}}\begin{sloppypar}\hyphenpenalty=5000\widowpenalty=500\clubpenalty=5000\input{B2G-16-003-authorlist.tex}\end{sloppypar}
\end{document}

%% file: B2G-16-003-authorlist.tex
\textbf{Yerevan Physics Institute,  Yerevan,  Armenia}\\*[0pt]
V.~Khachatryan, A.M.~Sirunyan, A.~Tumasyan
\vskip\cmsinstskip
\textbf{Institut f\"{u}r Hochenergiephysik,  Wien,  Austria}\\*[0pt]
W.~Adam, E.~Asilar, T.~Bergauer, J.~Brandstetter, E.~Brondolin, M.~Dragicevic, J.~Er\"{o}, M.~Flechl, M.~Friedl, R.~Fr\"{u}hwirth\cmsAuthorMark{1}, V.M.~Ghete, C.~Hartl, N.~H\"{o}rmann, J.~Hrubec, M.~Jeitler\cmsAuthorMark{1}, A.~K\"{o}nig, I.~Kr\"{a}tschmer, D.~Liko, T.~Matsushita, I.~Mikulec, D.~Rabady, N.~Rad, B.~Rahbaran, H.~Rohringer, J.~Schieck\cmsAuthorMark{1}, J.~Strauss, W.~Waltenberger, C.-E.~Wulz\cmsAuthorMark{1}
% ...
\vskip\cmsinstskip
% ...
1:~~Also at Vienna University of Technology, Vienna, Austria\\
% ...
